\documentclass[lettersize,journal]{IEEEtran}
\usepackage{cite}
\usepackage{bm}
\usepackage{amsmath,amsfonts}
\usepackage[noend]{algorithmic}
\usepackage{algorithm}
\usepackage{array}
\usepackage{subfig}
\usepackage{textcomp}
\usepackage{stfloats}
\usepackage{url}
\usepackage{verbatim}
\usepackage{graphicx}
\usepackage[font=small]{caption}  % 这里设置图注字体大小
\usepackage{booktabs} % 提供双横线功能
\usepackage{makecell}
\usepackage{multirow}
\usepackage{stfloats}
\usepackage{hyperref}

\usepackage{threeparttable}

\begin{document}

% \title{SITP: A Semantic Information Transport Protocol for Semantic Communication with TCP-Level Reliability and UDP-Level Latency}
\title{SITP: A High-Reliability Semantic Information Transport Protocol Without Retransmission \\ for Semantic Communication}

\author{
Yunhao Wang,~\IEEEmembership{Graduate Student Member,~IEEE,}
Shuai Ma,~\IEEEmembership{Member,~IEEE,} \\
Youlong Wu,~\IEEEmembership{Member,~IEEE,}
Guangming Shi,~\IEEEmembership{Fellow,~IEEE,}
Xiang Cheng,~\IEEEmembership{Fellow,~IEEE,} \\
Yuxuan Liu,
and Pengfei He,~\IEEEmembership{Graduate Student Member,~IEEE}

\thanks{This work was supported in part by the National Science and Technology Major Project-Mobile Information Networks under Grant No.2024ZD1300700, and in part by the Natural Science Foundation of China No.62293483. \textit{(Corresponding Author: Guangming Shi.)}}
\thanks{Yunhao Wang is with the School of Electronic and Computer Engineering, Peking University, Shenzhen 518055, China, and also with the Department of Networked Intelligence, Peng Cheng Laboratory, Shenzhen 518066, China (e-mail: yunhaowang@stu.pku.edu.cn)}
\thanks{Shuai Ma is with the Department of Networked Intelligence, Peng Cheng Laboratory, Shenzhen 518066, China. (e-mail: mash01@pcl.ac.cn).}
\thanks{Youlong Wu is with the School of Information Science and Technology, ShanghaiTech University, Shanghai 201210, China. (e-mail:wuyl1@shanghaitech.edu.cn).}
\thanks{Guangming Shi is with the Department of Networked Intelligence, Peng Cheng Laboratory, Shenzhen, 518066, China, and also with the School of Artificial Intelligence, Xidian University, Xi’an, Shaanxi 710071, China (e-mail: gmshi@pcl.ac.cn).}
\thanks{Xiang Cheng is with the State Key Laboratory of Photonics and Communications, School of Electronics, Peking University, Beijing 100871, China (e-mail: xiangcheng@pku.edu.cn).}
\thanks{Yuxuan Liu is with the School of Electronic and Computer Engineering, Shenzhen Graduate School, Peking University, Shenzhen 518055, China (e-mail: liuyuxuan@stu.pku.edu.cn).}
\thanks{Pengfei He is with the School of Artificial Intelligence, Xidian University, Xi’an, Shaanxi 710071, China, and also with the Department of Networked Intelligence, Peng Cheng Laboratory, Shenzhen 518066, China (e-mail: hepengfei@stu.xidian.edu.cn)}
\thanks{The code will be released on \url{https://github.com/WYHxuebi/SITP}.}
}

\maketitle

\begin{abstract}
With the evolution of 6G networks, modern communication systems are facing unprecedented demands for high reliability and low latency. However, conventional transport protocols are designed for bit-level reliability, failing to meet the semantic robustness requirements. To address this limitation, this paper proposes a novel Semantic Information Transport Protocol (SITP), which achieves TCP-level reliability and UDP-level latency by verifying only packet headers while retaining potentially corrupted payloads for semantic decoding. Building upon SITP, a cross-layer analytical model is established to quantify packet-loss probability across the physical, data-link, network, transport, and application layers. The model provides a unified probabilistic formulation linking signal noise rate (SNR) and packet-loss rate, offering theoretical foundation into end-to-end semantic transmission. Furthermore, a cross-image feature interleaving mechanism is developed to mitigate consecutive burst losses by redistributing semantic features across multiple correlated images, thereby enhancing robustness in burst-fade channels. Extensive experiments show that SITP offers lower latency than TCP with comparable reliability at low SNRs, while matching UDP-level latency and delivering superior reconstruction quality. In addition, the proposed cross-image semantic interleaving mechanism further demonstrates its effectiveness in mitigating degradation caused by bursty packet losses.
\end{abstract}


% \vfill\break

\section{Introduction}

\IEEEPARstart{W}{ithin} the IMT-2030 framework \cite{itu2023framework}, representative application scenarios—such as immersive extended reality (XR) \cite{alhakamy2024extended}, industrial IoT (IIoT) \cite{zhang2024time}, Vehicle-to-Everything (V2X) \cite{xu2024v2x}, and remote healthcare \cite{zha2024intelligent}—pose latency and reliability requirements that significantly surpass the capabilities of current wireless networks \cite{pennanen20246g}. These demands are now reflected in emerging application standards, which require millisecond-level or even sub-millisecond round-trip latency to support real-time interactions in human-centric and machine-centric communication systems \cite{tao20236g}. However, conventional bit-perfect communication protocols are fundamentally constrained by the latency–reliability trade-off: achieving low latency necessitates minimizing retransmissions and verification overhead, while attaining high reliability relies on multi-round control mechanisms that inevitably increase end-to-end delay \cite{li2024reliability}. This fundamental limitation has driven the evolution from bit-level reliability to semantic-level communication.

\begin{figure}[t]
    \centering
    \includegraphics[width=0.45\textwidth]{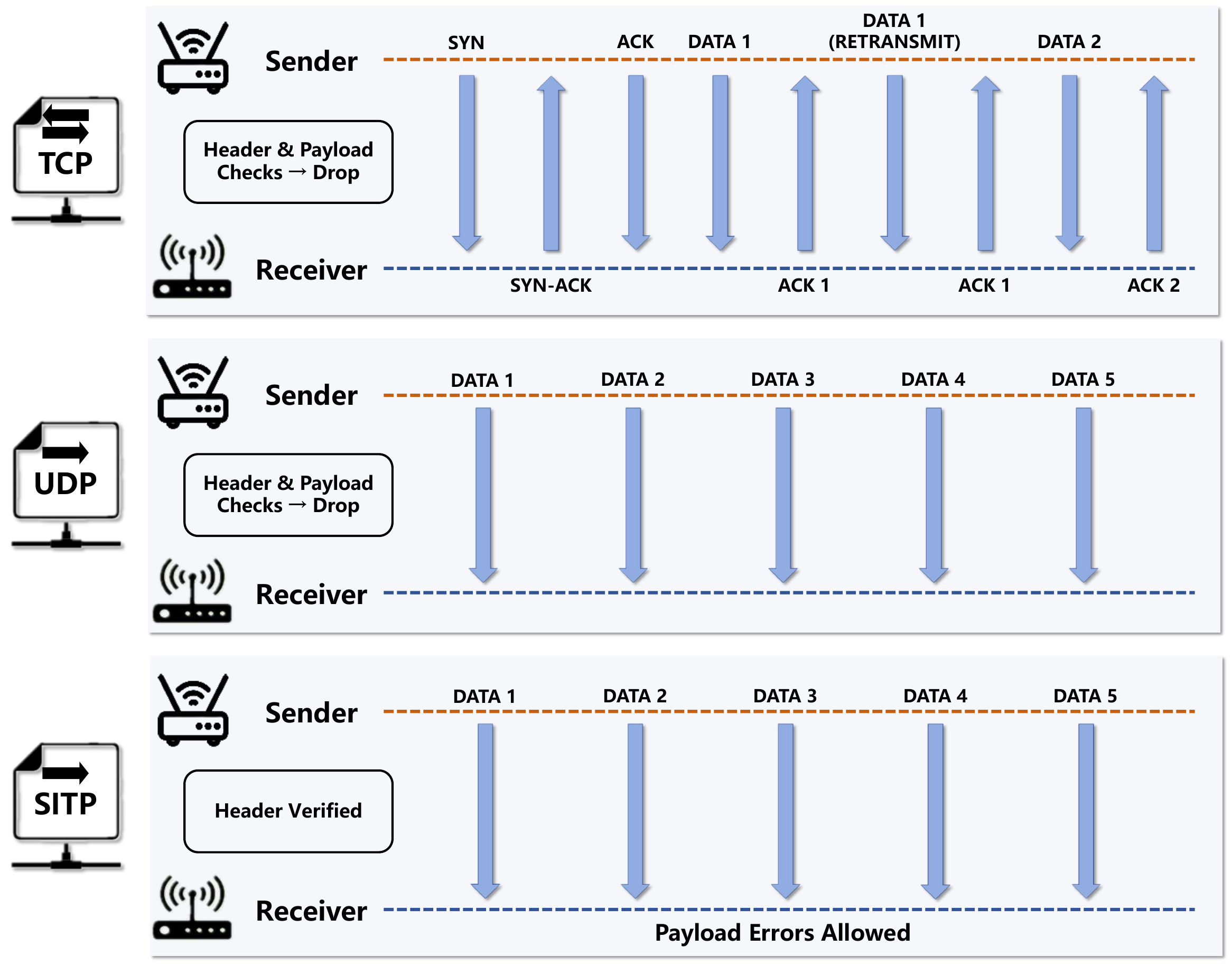}
    \caption{ The transport protocol mechanism comparison: TCP vs. UDP vs. SITP \textbf{(Ours)}. The proposed SITP validates only the header while tolerating residual noise in the payload.}
    \label{fig:ProtocolComparison}
\end{figure}

Semantic communication (SemCom) represents a paradigm shift from traditional bit-level transmission toward conveying task-relevant meaning \cite{shi2021new}. Unlike traditional systems focusing on bit-level accuracy, SemCom ensures that the intended meaning is successfully reconstructed at the receiver, even under partial corruption. Building upon Shannon and Weaver's notion of semantic-level communication, SemCom has recently attracted significant attention \cite{shannon1948mathematical, weaver1953recent}. With the rapid progress of deep learning (DL), encoder–decoder architectures enable semantic extraction, representation, and reconstruction, which demonstrates strong robustness: \textit{\textbf{even when semantic features subfigure are partially corrupted by noise, the receiver can still reconstruct approximate information that remains semantically consistent with the original intent.}}

Recent advances in SemCom have predominantly relied on joint source–channel coding (JSCC) frameworks, extending the applicability to multiple modalities such as text \cite{mao2024gan}, speech \cite{weng2023deep}, image \cite{wang2024swin, shen2025semantic}, video \cite{teng2025conquering}, and 3D point clouds \cite{chen2024cross}. \textit{\textbf{However, most existing research remains confined to a physical-layer abstraction, typically assuming additive or fading channels while overlooking the packetization and protocol-layer mechanisms inherent in real systems.}} In practical network, transmitted data are segmented, encapsulated, and verified across multiple protocol layers, where packet loss-rather than bit error, typically constitutes the primary source of semantic degradation. Consequently, current research fails to capture the end-to-end characteristics of realistic digital systems. A few studies have attempted to model semantic erasure channels, typically assuming independent packet losses \cite{tian2025synchronous, teng2025conquering}. \textit{\textbf{However, under continuous burst interference, packet headers or fragments become undecodable, resulting in consecutive losses and severe semantic distortion.}} Existing intra-image interleaving remains insufficient to prevent semantic collapse under burst losses.

\captionsetup[table]{justification=centering, labelsep=space, textfont=sc} % 设置标题全大写
\begin{table*}[t]
    \renewcommand\arraystretch{2.3}
    \setlength{\tabcolsep}{8pt}
    \caption{ \\ Contrasting our Contributions to the Transport Protocols \label{tab1}}
    \centering
    \begin{tabular}{ccccccc}
        \hline
        \hline
        \textbf{Protocol} & \textbf{Handshake \& Retrans.} & \makecell{\textbf{Transport-Layer} \\ \textbf{Validation Coverage}} & \makecell{\textbf{Data-Link Layer} \\ \textbf{CRC Coverage}} & \textbf{Reliability} & \textbf{Latency} & \textbf{Semantic Suitability}\\
        \hline
        TCP \cite{feng2025exploiting} & \makecell{3-way handshake \\ and ACK retrans.} & Header + Payload & Header + Payload & High & High & \multirow{2}{*}{\makecell{Payload checksum failure \\ causes packet discard, \\ making  payloads unusable.}}\\
        UDP \cite{jiang2025position} & \makecell{No handshake \\ or retrans.}  & Header + payload & Header + Payload & Low & Low \\
        \textbf{SITP (Ours)} & \makecell{\textbf{No handshake} \\ \textbf{or retrans.}}  & \textbf{Header-only} & \textbf{Header-only} & \textbf{High} & \textbf{Low} & \makecell{\textbf{Noisy payload retained} \\ \textbf{for semantic decoding.}} \\
        \hline
        \hline
    \end{tabular}
    \begin{tablenotes}
        \footnotesize
        \item[*] \text{Noting:} \textbf{Retrans.} denotes Retransmissions. The data-link layer CRC covers only the header, ensuring that payload corruption does not trigger packet drops.
    \end{tablenotes}
\end{table*}

From a cross-layer perspective, current communication primarily rely on Transmission Control Protocol (TCP) \cite{feng2025exploiting} or User Datagram Protocol (UDP) \cite{jiang2025position} at the transport layer. \textit{\textbf{However, both protocols were fundamentally designed to ensure bit-level reliability rather than semantic robustness, rendering them inadequate for exploiting the intrinsic resilience of SemCom to channel impairments.}} Specifically, TCP achieves high reliability through acknowledgment and retransmission mechanisms, which introduce considerable latency, making it unsuitable for delay-sensitive scenarios. In contrast, UDP provides low-latency transmission without retransmission but discards entire packets upon checksum errors, thereby wasting payloads that still contain semantically valuable information. Such packet-level discard behavior embodies the conventional design paradigm of ensuring bit-perfect accuracy, which fundamentally conflicts with the principle of SemCom—preserving the conveyed meaning rather than guaranteeing error-free bits.

In summary, existing research still faces several fundamental challenges that constrain the advancement of SemCom:
\begin{itemize}
\item{A transport-layer protocol tailored to SemCom is required to integrate UDP's low latency  with TCP's high reliability, which exploits the semantic value of imperfect payloads rather than discarding them upon error detection.}
\item{A robust mechanism is essential to enhance resilience against consecutive packet losses, thus preserving the semantic feature of transmitted images.}
\end{itemize}

\subsection{Contributions}

To the best of our knowledge, there has been no study that modifies transport-layer protocol mechanisms to retain interference-affected payloads carrying semantically meaningful information, thereby enhancing semantic-level robustness. The main contributions are summarized as follows:

\begin{itemize}
\item{\textbf{Semantic Information Transport Protocol (SITP):} We propose a novel transport-layer protocol, termed Semantic Information Transport Protocol. Unlike TCP, the proposed SITP eliminates the three-way handshake and retransmission mechanisms, thereby effectively reducing end-to-end latency. Compared with UDP, SITP tolerates bit errors within data segments rather than discarding corrupted packets, which enabling the receiver to leverage the residual semantic information, thereby enhancing the reliability of transmission.}
\item{\textbf{Cross-layer mathematical model of packet loss:} Based on SITP, a cross-layer mathematical model is established to characterize the packet-loss probability across the physical, data link, network, transport, and application layers. Furthermore, the SNR and packet-loss rate are integrated into a unified analytical formulation, enabling systematic analysis of transmission performance.}
\item{\textbf{Cross-images feature-level interleaving:} Building on SITP, we further design a SemCom system employing a cross-image feature-level interleaving mechanism. By distributing lost semantic features across multiple correlated images, the proposed approach significantly enhances system robustness against burst interference and consecutive packet losses.}
\item{\textbf{Extensive experiments}  demonstrate that SITP achieves lower latency than TCP while maintaining comparable performance. Compared with UDP, SITP attains similar transmission latency yet delivers significantly better reconstruction quality. Moreover, the proposed cross-image semantic interleaving substantially enhances robustness against consecutive packet losses.}
\end{itemize}

\subsection{Organization of This Article}

The rest of this paper is organized: Section II reviews related work on SemCom, transport-layer protocols, and interleaving techniques. Section III presents the system model and the proposed SITP. Section IV formulates a cross-layer packet-loss model based on SITP. Section V details a image transmission SemCom system with cross-image feature-level interleaving for burst-loss resilience. Section VI provides simulation results and analysis, and Section VII concludes the paper.

\section{Related work}

\subsection{Transport-Layer Protocols}

At the transport layer, existing architectures rely primarily on the classical protocols—TCP and UDP \cite{haryono2025comparative}. TCP, as a connection-oriented protocol, ensures error-free delivery through mechanisms such as the handshake, acknowledgments, and automatic repeat request (ACK) \cite{feng2025exploiting}. Although the mechanisms ensure bit-level reliability, they introduce substantial latency, making TCP unsuitable for latency-sensitive applications. In contrast, UDP provides a connectionless datagram service with minimal overhead \cite{jiang2025position}. The checksum-based error detection verifies the integrity of both the header and payload, discarding any packet failing validation, which achieves low latency but sacrifices reliability.

Beyond these traditional schemes, several alternative transport protocols have been proposed \cite{10.1145/2333112.2333113, amend2019framework, michel2022flec, kutsevol2023towards}. A Stream Control Transmission Protocol (SCTP) framework was introduced in \cite{10.1145/2333112.2333113}, providing message-oriented, multistream transmission with built-in congestion and reliability control. In \cite{amend2019framework}, a multipath Datagram Congestion Control Protocol (DCCP) framework was introduced to support IP-compatible multiaccess transmission through adaptive packet scheduling. \cite{michel2022flec} proposed a QUIC-based protocol featuring flexible frame-level encoding and application-tailored reliability mechanisms to enhance end-to-end efficiency. A semantic-aware transport policy was introduced in \cite{kutsevol2023towards}, which relies on ACK-based retransmission and jointly optimizes packet admission and update triggering according to application-level importance. Nevertheless, the fundamental paradigm remains unchanged: existing protocols still enforce payload integrity through checksums or retransmission, such that any detected corruption in packet discarding. To bridge this gap, the proposed SITP introduces a novel paradigm that preserves and leverages degraded payloads carrying meaningful semantic information, thereby enhancing reliability without increasing latency.

\subsection{Semantic Communication}

In the field of SemCom, extensive studies have explored JSCC-based frameworks under additive white Gaussian noise (AWGN) and various fading channels, demonstrating the ability of SemCom systems to effectively tolerate channel impairments \cite{mao2024gan, weng2023deep, wang2024swin, shen2025semantic, teng2025conquering, chen2024cross}. A GAN-based SemCom named Ti-GSC was proposed in \cite{mao2024gan}, enabling text transmission over fading channels without requiring CSI through a GAN-assisted distortion suppression module. In \cite{weng2023deep}, DeepSC-ST jointly performs speech recognition and synthesis by transmitting text-related semantic features, achieving robust performance. The author in \cite{chen2024cross} proposed a cross-modal graph SemCom assisted by generative AI, integrating graph neural networks to enable robust multimodal transmission. However, these studies are predominantly confined to physical-layer abstractions, where only channel impairments are modeled, while practical aspects such as packet segmentation, header processing, and payload verification are largely ignored.

Research on packet-loss scenarios remains limited, with only a few studies attempting to model the erasure channels. A packet-loss-resistant video SemCom system named MSTVSC was proposed in \cite{teng2025conquering}, employing a MoE Swin Transformer and packet-loss recovery to maintain high-quality reconstruction. \cite{tian2025synchronous} proposed a synchronous multi-modal SemCom system, integrating packet-level forward-error-correction and cross-modal alignment to achieve semantics-synchronized transmission. However, these studies still adhere to conventional transport protocols, without leveraging the potential semantic information embedded in noisy payloads. This paper builds upon the proposed SITP to establish a cross-layer mathematical model that jointly considers the SNR and packet-loss rate, thereby enabling a practical SemCom architecture.

\subsection{Semantic Interleaving Mechanisms}

Current interleaving techniques operate at the bit or symbol level by rearranging the encoded sequence prior to transmission to mitigate burst errors. In SemCom, only a limited number of studies have extended interleaving to the image-feature or packet level. In \cite{teng2025conquering}, application-layer semantic-level interleaving was proposed to disperse correlated semantic features and mitigate concentrated semantic loss caused by packet drops. However, as these methods remain confined to spatial rearrangement within a single image, consecutive packet losses still corrupt global semantic features. This paper proposes a cross-image semantic interleaving mechanism that redistributes semantic representations across multiple images, allowing semantic information to be jointly preserved across frames and thereby enhancing transmission robustness.

% \textbf{Notation}: matrices, vectors and scalars are represented by bold uppercase letters, bold lowercase letters and lowercase letters, respectively. The sets of real and complex numbers are represented by \(\mathbb{R}\) and \(\mathbb{C}\), respectively.

\section{Semantic Information Transport Protocol}

\begin{figure*}[t]
\centering
\includegraphics[width=0.98\textwidth]{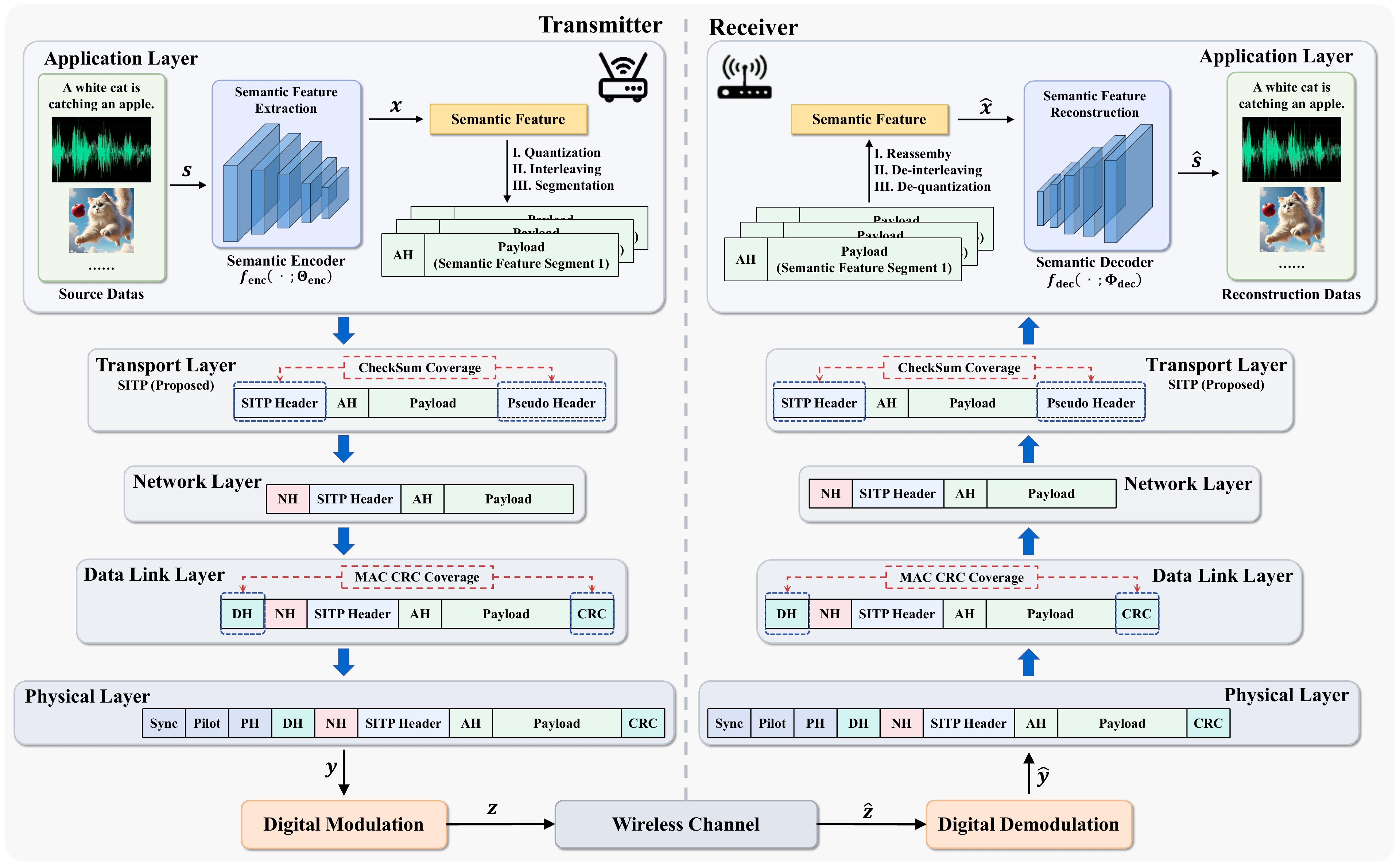}
\caption{The cross-layer architecture of digital semantic communication based on SITP. The SITP-based framework integrates cross-layer SemCom, verifying only headers while preserving noisy payloads for semantic reconstruction. Note: AH, NH, DH, and PH denote the application-layer header, network header, data-link header, and physical-layer header, respectively.}
\label{fig:Architecture}
\end{figure*}

As illustrated in Fig.\ref{fig:Architecture}, the proposed cross-layer digital SemCom adopts the general layered architecture of modern network, consisting of the application, transport, network, data-link, and physical layers, while excluding the presentation and session layers for simplicity in this work.

\subsection{System Model}

At the transmitter, the source data $\bm{s}$ (e.g., text, speech, image, or video) is processed by a semantic encoder to extract the latent semantic representation $\bm{x}$. Formally, the encoding process can be expressed as:
\begin{equation}
    \bm{x} = f_{\text{enc}}(\bm{s} \, ; \, \bm{\Theta_{\textbf{enc}}}),
    \label{eq:enc}
\end{equation}
where $f_{\text{enc}}(\, \bm{\cdot} \,;\, \bm{\Theta_{\textbf{enc}}} \,)$ denotes the semantic encoding function parameterized by the trainable network parameters $\bm{\Theta_{\textbf{enc}}}$. 

The extracted semantic features $\bm{x}$ are quantized into bitstreams and subsequently processed by a semantic-level interleaver, which redistributes the feature elements. Details of the semantic-level interleaving mechanism are provided in Section IV. The interleaving process can be expressed as:
\begin{equation}
\bm{k} = g_{\text{int}}(\mathcal{Q}(\bm{x}) \, ; \, \bm{\Pi}),
\label{eq:intlv}
\end{equation}
where $g_{\text{int}}(\, \bm{\cdot} \, ; \, \bm{\Pi})$ denotes the interleaving function, $\bm{\Pi}$ denotes the semantic interleaving index vector, representing the permutation order of feature elements, $\bm{k}$ denotes the interleaved feature vector, and $\mathcal{Q}(\cdot)$ denotes the quantization function.

After interleaving, the semantic features are packetized at the application layer, where an application-layer header (AH) is appended. The payload, together with the AH, is then delivered to the transport layer, where the proposed Semantic Information Transport Protocol (SITP) encapsulates the segment with its own header and Pseudo-Header. The detailed design of the SITP header format and checksum verification mechanism will be presented in Section III-B. The encapsulated segment is subsequently passed to the network layer for IP-level framing, where a network header (NH) is appended. At the data-link layer, a data-link header (DH) and a cyclic redundancy check (CRC) code are futher attached. \textit{\textbf{Unlike conventional systems, where the CRC verification covers both the DH and the payload, the proposed framework modifies the CRC coverage to include only the header portion.}} Such modification is consistent with the proposed SITP protocol, as the payload may contain noisy yet semantically valuable features that should not trigger packet discarding during error detection. Finally, at the physical layer, synchronization sequences (Sync), pilot symbols (Pilot), and physical headers (PH) are appended, forming a sequence of transmitted packets denoted as $\bm{y} \in \mathbb{R}^{T \times L}$, where $T$ represents the total number of packets and $L$ denotes the length of each packet. Formally, the cross-layer packetization processes can be expressed as:
\begin{equation}
\bm{y} = h_{\text{pkt}}(\bm{k}),
\label{eq:packetization}
\end{equation}
$h_{\text{pkt}}(\cdot)$ represents the cross-layer packetization function.

\begin{figure*}[t]
\centering
\includegraphics[width=0.90\textwidth]{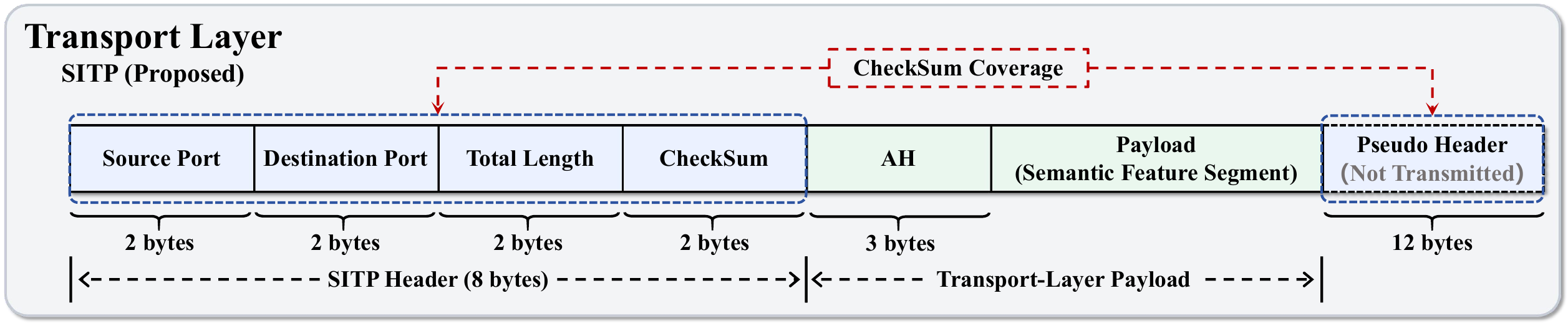}
\caption{The overall structure of the proposed SITP packet at the transport layer. The checksum operation verifies only the SITP header fields while excluding the AH and payload segments, allowing noisy payloads to be received and utilized for semantic reconstruction.}
\label{fig:SITPstructure}
\end{figure*}   

After packetization, the segmented data $\bm{y}$ are modulated to generate the constellation symbols $\bm{z}$. The digital modulation process can be expressed as
\begin{equation}
\bm{z} = \mathcal{M}_{\text{mod}}(\bm{y} \, ; \, M),
\label{eq:mod}
\end{equation}
where $\mathcal{M}(\cdot  \, ; \, M)$ represents the modulation function, $M$ denotes the modulation order (e.g., $M$-QAM), and $\bm{z} \in \mathbb{C}^{T \times L/log_{2}M}$ denotes the constellation symbols. In practical systems, headers typically employ more robust, lower-rate modulation for reliability, whereas payloads use higher-rate schemes for efficiency. For simplicity, identical modulation is applied to both in this work.

Through the wireless channel, the received signal $\bm{\hat{z}} \in \mathbb{C}^{T \times L/log_{2}M}$ can be expressed as
\begin{equation}
    \bm{\hat{z}} = \sqrt{P} \bm{H} \bm{z} + \bm{n},
    \label{eq:channelpass}
\end{equation}
% where $\bm{H}$ is the channel matrix, $\bm{n} \sim \mathcal{CN}(0, \sigma_n^2 \bm{I_{M_t}})$ is the additive white Gaussian noise, and $P$ is the transmit power.
where $\bm{H}$ is the channel matrix, $\bm{n}$ is the additive white Gaussian noise, and $P$ is the transmit power.

\textbf{Noting:} Although the wireless channel in this study still follows \eqref{eq:channelpass} commonly adopted in SemCom research, the proposed framework further considers the effect on protocol headers. During cross-layer depacketization, header corruption may cause parsing failures and resulting packet losses, which provides a more accurate representation of end-to-end semantic transmission in practical systems.

The receiver adopts a symmetric architecture with respect to the transmitter. The received signal $\bm{\hat{z}}$ is sequentially demodulated and decapsulated through the multiple layers, where each layer independently verifies its corresponding header integrity (e.g., via CRC or checksum). At the application layer, the correctly received packets are reassembled and processed through de-interleaving and dequantization to reconstruct the original semantic feature $\bm{\hat{x}}$. Formally, the recovery process can be expressed as
\begin{equation}
    \bm{\hat{y}} = \mathcal{M}_{\text{demod}}(\bm{\hat{z}} \, ; \, M), 
    \label{eq:demod}
\end{equation}
\begin{equation}
    \bm{\hat{k}} = h_{\text{depkt}}(\bm{\hat{y}}), 
    \label{eq:unpkt}
\end{equation}
\begin{equation}
    \bm{\hat{x}} = \mathcal{Q}^{-1}(g_{\text{deint}}(\bm{\hat{k}} \, ; \, \bm{\Pi})),  
    \label{eq:unintlv}
\end{equation}
where $\mathcal{M}_{\text{demod}}(\cdot)$ denotes the digital demodulation function, $g_{\text{depack}}(\cdot)$ denotes the depacketization function, and $g_{\text{deint}}(\cdot,;\bm{\Pi})$ denotes the de-interleaving operation that restores the original semantic feature order using the interleaving index matrix $\bm{\Pi}$, and $\mathcal{Q}^{-1}(\cdot)$ denotes the de-quantization function. Finally, the reconstructed semantic features are decoded by the semantic decoder to recover the estimated source data $\bm{\hat{s}}$:
\begin{equation}
\bm{\hat{s}} = f_{\text{dec}}(\bm{\hat{x}} \, ; \, \bm{\Theta}_{\textbf{dec}}),
\label{eq:dec}
\end{equation}
where $f_{\text{dec}}(\cdot \, ; \, \bm{\Theta}_{\textbf{dec}})$ is the semantic decoding function parameterized by $\bm{\Theta}_{\textbf{dec}}$. 

\subsection{Protocol Design and Header Structure}

As illustrated in Fig.\ref{fig:SITPstructure}, we introduce the Semantic Information Transport Protocol (SITP) and specify its packet format and checksum policy. SITP is designed to achieve UDP-class latency by eliminating connection setup and retransmission, while simultaneously approaching TCP-level reliability through the delivery of payloads that remain semantically meaningful even when partially corrupted. The SITP  explicitly leverages the error tolerance of SemCom while avoiding the feedback delays inherent in traditional transport schemes.

\captionsetup[table]{justification=centering, labelsep=space, textfont=sc} % 设置标题全大写
\begin{table}
    \renewcommand\arraystretch{1.8}
    \setlength{\tabcolsep}{6pt}
    \caption{ \\ SITP Header Format (8 bytes, fixed length) \label{tab:SITPHeader}}
    \centering
    \begin{tabular}{ccc}
        \hline
        \hline
        \textbf{Field} & \textbf{Length (bytes)} & \textbf{Description} \\
        \hline
        Source Port & 2 & Sending port identifier \\
        Destination Port & 2 & Receiving port identifier \\
        Total Length & 2 & TH + Transport payload \\
        CheckSum & 2 & \makecell{Header checksum \\ (with pseudo-header)}\\
        \hline
        \hline
    \end{tabular}
\end{table}

Unlike conventional bit-perfect transport protocols, the SITP performs integrity verification exclusively on the header fields rather than the entire segment. The checksum computation covers the SITP header together with a pseudo-header that replicates essential network-layer and transport-layer context to prevent misrouting or misdelivery. The pseudo-header is not transmitted, which serves only as auxiliary information for checksum \cite{degermark1996low}.

By restricting verification to the head fields, SITP ensures correct packet identification while omitting payload checks, thereby preventing unnecessary integrity validation on semantic feature. When the header verification fails, the corresponding packet is discarded at the receiver. In contrast, if the header passes integrity verification, the associated payload, even when affected by bit errors, is retained and forwarded to the application layer. The semantic decoder reconstructs the intended meaning of the received data, enabling the system to maintain reliable performance even under imperfect transmission conditions. Formally, the checksum generation and acceptance rules can be expressed as
\begin{equation}
    \text{CS} = \text{CheckSum}(\textbf{SITP\_HDR}\,\Vert\,\textbf{Pseudo\_HDR}),
    \label{eq:checksum}
\end{equation}
where $\Vert$ denotes the concatenation operator, $\text{CheckSum}(\cdot)$ represents a one’s-complement checksum \cite{braden1989computing}, $\textbf{SITP\_HDR}$ refers to the SITP header, and $\textbf{Pseudo\_HDR}$ denotes the pseudo header. A checksum result of $\mathrm{CS}=0$ indicates a valid header, while any nonzero value denotes verification failure.

As summarized in Table \ref{tab:SITPHeader}, the SITP header follows a 8-byte design similar to the UDP header, consisting of four fundamental fields: Source Port, Destination Port, Total Length, and Checksum, each occupying two bytes \cite{jiang2025position}. The Total Length field specifies the overall packet size, including the SITP header, the AH, and the application payload. In contrast, TCP \cite{feng2025exploiting} employs a variable-length header ranging from 20 to 60 bytes depending on optional fields.

Accordingly, SITP delivers all packets whose headers pass integrity verification, which enables the upper application layer to utilize the residual semantics embedded in noisy payloads, thereby achieving UDP-level latency without retransmission while maintaining TCP-level reliability.

\textbf{Noting:} The proposed SITP framework is subject to two types of distortion in semantic transmission. Specifically, when the SITP header fails checksum verification, the corresponding packet is discarded. Conversely, when the header passes verification, the payload remains susceptible to noise introduced by the physical channel. Therefore, the SITP-based system experiences both packet-level and feature-level impairments.

\section{Mathematical Modeling of Cross-Layer Packet Loss in SITP}

In this section, a comprehensive mathematical framework is developed to model the packet loss probability of the proposed SITP across multiple layers, including the Physical Layer, Data Link Layer, Network Layer, Transport Layer, and Application Layer. Each layer’s packet loss characteristics are independently modeled according to its mechanisms. Subsequently, a cross-layer coupling model is derived to capture the compound effects of bit errors, synchronization failures, and checksum errors, thereby providing a more realistic characterization of SITP performance in practical SemCom systems.

At the physical layer, a transmitted packet primarily consists of a synchronization sequence, pilot symbols, the PH, and the physical-layer payload. In this work, synchronization is assumed to be achieved through a correlation-based detection. Consequently, a packet is considered lost if the number of bit errors within the synchronization sequence exceeds a predefined tolerable threshold ${t_\text{sync}}$. Therefore, the successful synchronization probability $P_{\text{sync-suc}}$ can be expressed as a binomial cumulative distribution:
\begin{equation}
    P_{\text{sync-suc}} = \sum_{i=0}^{t_\text{sync}} \tbinom{8 \cdot N_\text{sync}}{i} P_{b}^{i}(1-P_{b})^{8 \cdot N_\text{sync}-i},
    \label{eq:syncsuc}
\end{equation}
where $P_{b}$ denotes the bit error rate (BER), and ${N_\text{sync}}$ denotes the total length (bytes) of the frame synchronization sequence.

Once synchronization is achieved, the correctness of the PH determines whether the packet can be successfully decoded. The header detection success probability $P_{\text{Phy-suc}}$ can be expressed as:
\begin{equation}
    P_{\text{PH-suc}} = (1-P_{b})^{8 \cdot N_\text{PH}},
    \label{eq:phsuc}
\end{equation}
where ${N_\text{PH}}$ denotes the length of the PH (bytes). The physical-layer packet loss probability $P_{\text{Phy-fail}}$ can be expressed as:
\begin{align}
     P_{\text{Phy-fail}} & = 1-P_{\text{sync-suc}} \cdot P_{\text{PH-suc}} 
     \label{eq:phyfail1} \\
    & = 1 - [\sum_{i=0}^{t_\text{sync}} \tbinom{8 \cdot N_\text{sync}}{i} P_{b}^{i}(1-P_{b})^{8 \cdot N_\text{sync}-i}] 
    \notag \\
     & \cdot (1-P_{b})^{8 \cdot N_\text{PH}}.
     \label{eq:phyfail2}
\end{align}

At the data link layer, as illustrated in Fig.\ref{fig:Architecture}, the proposed SITP redefines the cyclic redundancy check (CRC) coverage range compared with traditional schemes. Specifically, the CRC verification is applied only to the DH and the CRC field, while the data-link-layer payload is excluded from the error detection process. Assuming that the data link layer employs an $r_{d}$-bit CRC code for error detection. In this case, two conditions result in successful packet reception: (I) the DH is error-free, or (II) bit errors occur in the DH but remain undetected by the CRC. Accordingly, the successful transmission probability $P_{\text{Dalink-suc}}$ of the data link layer can be expressed as \cite{lin2005fast}:
\begin{align}
     P_{\text{Dalink-suc}} & = P_{\text{DH}} + (1 - P_{\text{DH}}) \cdot 2^{-r_{d}} 
     \label{eq:dalinksuc1} \\
     & = (1 - P_{b})^{8 N_{\text{DH}}}+ (1 - (1 - P_{b})^{8 N_{\text{DH}}}) \cdot 2^{-r_{d}} \label{eq:dalinksuc2} \\
     & = 1 - [1 - (1 - P_{b})^{8 \cdot N_{\text{DH}}}] \cdot (1 - 2^{-r_{d}}), 
     \label{eq:dalinksuc3}
\end{align}
where $P_{\text{DH}}$ denotes the probability that the DH is received without error, and ${N_\text{DH}}$ denotes the length of the DH (bytes). Thus, the corresponding packet loss probability $P_{\text{Dalink-fail}}$ at the data link layer is:
\begin{equation}
    P_{\text{Dalink-fail}} = [1 - (1 - P_{b})^{8 \cdot N_{\text{DH}}}] \cdot (1 - 2^{-r_{d}}).
    \label{eq:dalinkfail}
\end{equation}

At the network layer, each SITP packet is encapsulated within the IP datagram. Any bit error occurring within the NH results in the packet being discarded. Thus, the network-layer packet loss probability $P_{\text{Net-fail}}$ can be expressed as:
\begin{equation}
    P_{\text{Net-fail}} = 1 - (1 - P_{b})^{8 \cdot N_{\text{NH}}}, 
    \label{eq:netkfail}
\end{equation}
where ${N_\text{NH}}$ denotes the length of the NH (bytes).

At the transport layer, the proposed SITP is employed. As illustrated in Fig.\ref{fig:Architecture}, the checksum coverage of SITP is computed over the SITP header and a pseudo header. Similar to the data-link layer, a packet at the transport layer is considered lost if the SITP header is corrupted and the checksum successfully detects the error. According to \cite{partridge1995performance}, the undetected-error probability of an $r_{s}$-bit checksum can be approximated by $2^{-r_{s}}$ under the random independent bit-error assumption. Accordingly, the packet loss probability $P_{\text{SITP-fail}}$ of the transport layer is given by
\begin{equation}
    P_{\text{SITP-fail}} = [1 - (1 - P_{b})^{8 \cdot N_{\text{SITP\_HDR}}}] \cdot (1 - 2^{-r_{s}}), 
    \label{eq:SITPfail}
\end{equation}
where ${N_\text{SITP\_HDR}}$ denotes the length of the SITP header (bytes). It should be noted that the pseudo header, while included in the checksum computation, is not transmitted over the channel.

\begin{figure*}[t]
\centering
\includegraphics[width=\textwidth]{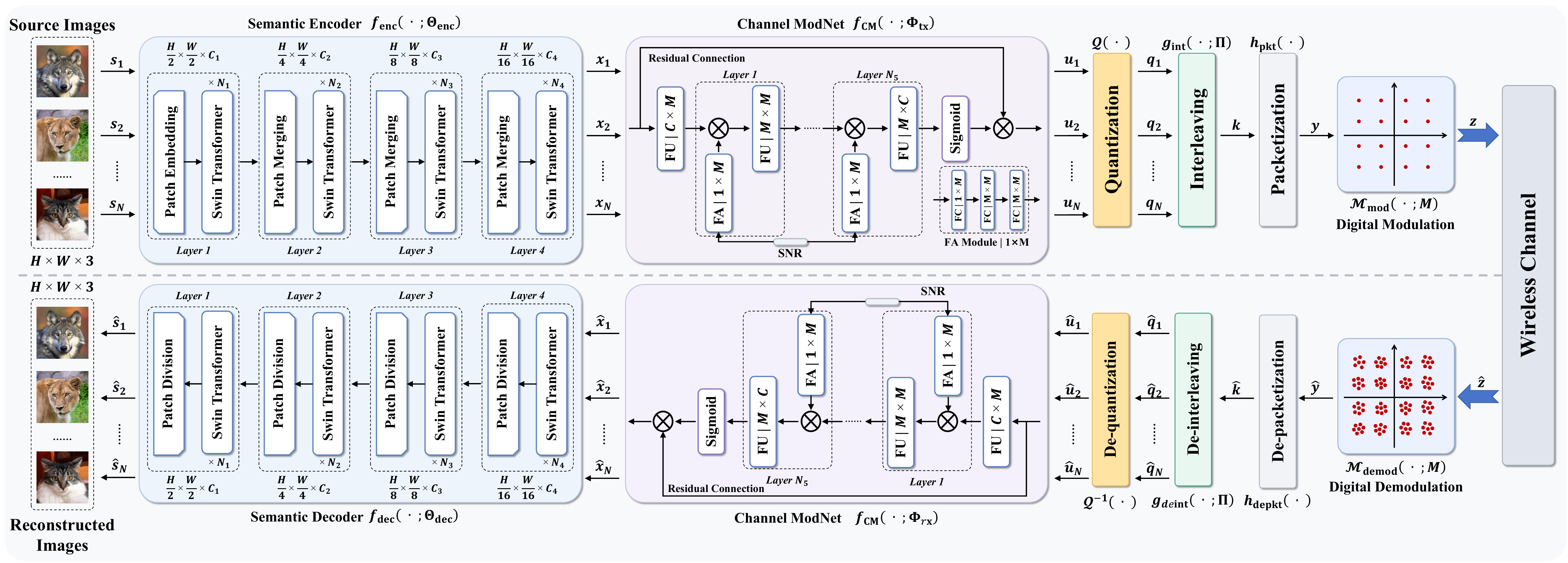}
\caption{The overall architecture of the proposed SITP-based digital semantic communication system for the burst-loss resilience.}
\label{fig:overallArchitecture}
\end{figure*}

At the application layer, each transmitted semantic feature segment is assigned an AH, following the structure proposed in \cite{teng2025conquering}. Successful decoding is ensured only when the corresponding application header is received error-free. Therefore, the packet loss probability $P_{\text{App-fail}}$ at the application layer can be expressed as
\begin{equation}
    P_{\text{App-fail}} = 1 - (1 - P_{b})^{8 \cdot N_{\text{AH}}}, 
    \label{eq:appfail}
\end{equation}
where ${N_\text{AH}}$ denotes the length of the AH (bytes).

Based on the models derived in \eqref{eq:phyfail2}, \eqref{eq:dalinkfail}-\eqref{eq:appfail}, the packet loss probabilities of each layer have been independently formulated. Assuming that packet loss across different layers are statistically independent, the overall probability that a packet is successfully delivered across the entire communication stack can be expressed as \eqref{eq:crossfail2}, shown at the bottom of this page.

\begin{figure*}[b]
    \hrulefill
    \centering
    \begin{align}
        P_{\text{Cross-fail}}(P_{b}) & = 1 - (1 - P_{\text{Phy-fail}})(1 - P_{\text{Dalink-fail}})(1 - P_{\text{Net-fail}})(1 - P_{\text{SITP-fail}})(1 - P_{\text{App-fail}}),
        \label{eq:crossfail1} \\
         & = 1 - (1 - P_{b})^{8 \cdot (N_{\text{PH}} + N_{\text{NH}} + N_{\text{AH}})} \cdot [\sum_{i=0}^{t_\text{sync}} \tbinom{8 \cdot N_\text{sync}}{i} P_{b}^{i}(1-P_{b})^{8 \cdot N_\text{sync}-i}] \cdot \{1 - [1 - (1 - P_{b})^{8 \cdot N_{\text{DH}}}] \cdot (1 - 2^{-r_{d}})\} 
        \notag \\
         & \cdot \{1 - [1 - (1 - P_{b})^{8 \cdot N_{\text{SITP\_HDR}}}] \cdot (1 - 2^{-r_{s}})\}
         \label{eq:crossfail2}
    \end{align}
\end{figure*}

From the derived cross-layer formulation, it can be observed that, under a given configuration, the parameters $N_{\text{PH}}$, $N_{\text{NH}}$, $N_{\text{AH}}$, $N_{\text{DH}}$, $N_{\text{SITP\_HDR}}$, $N_{\text{sync}}$, $t_{\text{sync}}$, $r_{d}$ and $r_{s}$ are all predetermined by the protocol design. Consequently, the overall cross-layer packet loss rate in SITP becomes exclusively determined by the underlying BER, distinguishing it from conventional TCP/UDP frameworks, in which the packet loss probability is explicitly dependent on the packet length. \textit{\textbf{Hence, SITP’s end-to-end reliability depends on channel conditions rather than on the size of transmitted payloads.}}

We further consider a digital SemCom system employing QAM with ideal channel equalization. Under the assumptions, the received signal can be modeled as the AWGN channel. According to \cite{raju2016evaluation}, the theoretical relationship between the SNR and the BER for an $M$-QAM is given by
\begin{equation}
    P_{b}( \gamma_{b}) \approx \frac{4}{log_{2}M}(1-\frac{1}{\sqrt{M}})Q(\sqrt{\frac{3log_{2}M \cdot \gamma_{b}}{M-1}}), 
    \label{eq:qam}
\end{equation}
where $Q(\cdot)$ is the Gaussian $Q$-function, and $\gamma_{b}$ denotes the bit-level SNR, defined as 
\begin{equation}
    \gamma_{b} = \frac{E_{b}}{N_{0}} = \frac{E_{s}}{N_{0} \cdot log_{2}M}.
    \label{eq:snr}
\end{equation}

Therefore, based on \eqref{eq:crossfail2}–\eqref{eq:snr}, a cross-layer mathematical model of packet loss is established for the SITP-based transmission framework. The model captures the relationship between the signal-to-noise ratio (SNR) and the overall packet loss probability across the entire SemCom stack, thereby enabling a systematic evaluation of transmission reliability.

\section{SITP-Based Semantic Communication System for Burst-Loss Resilience}

In this section, based on the SITP transmission protocol, we present a digital SemCom system for image transmission under consecutive burst packet losses. Inspired by \cite{yang2024swinjscc}, the SwinJSCC is adopted as both the semantic encoder and decoder. Furthermore, a cross-image feature interleaving mechanism is incorporated to enhance robustness against burst losses.

\subsection{The Overall Architecture}

The overall framework is illustrated in Fig.\ref{fig:overallArchitecture}, where the transmitter consists of several components, including the Semantic Encoder, Channel ModNet, Quantization, Interleaving, Packing, and Digital Modulation. 

In the semantic encoder, the Patch Embedding layer divides each input image $\bm{s_i} \in \mathbb{R}^{H \times W \times 3}$ into non-overlapping patches of size $\frac{H}{2} \times \frac{W}{2}$. Then the Swin Transformer Block processes these features while maintaining the same resolution. To extract deeper semantic representations, the Patch Merging layer aggregates neighboring features and performs down-sampling through linear projection, thereby halving the number of tokens \cite{yang2024swinjscc}.  The semantic representation is given by:
\begin{equation}
    \bm{x_i} = f_{\text{enc}}(\bm{s_i} \, ; \, \bm{\Theta_{\textbf{enc}}}), \,\,\,\, \bm{x_i} \in \mathbb{R}^{M_t \times 1},
    \label{eq:encoderimgs}
\end{equation}
where $\bm{x_i}$ denotes the semantic feature of image $i$.

\textbf{Noting:} As the cross-image semantic feature interleaving mechanism is employed, semantic encoding must be performed on multiple images.

\begin{figure*}[t]
    \centering
    \includegraphics[width=\textwidth]{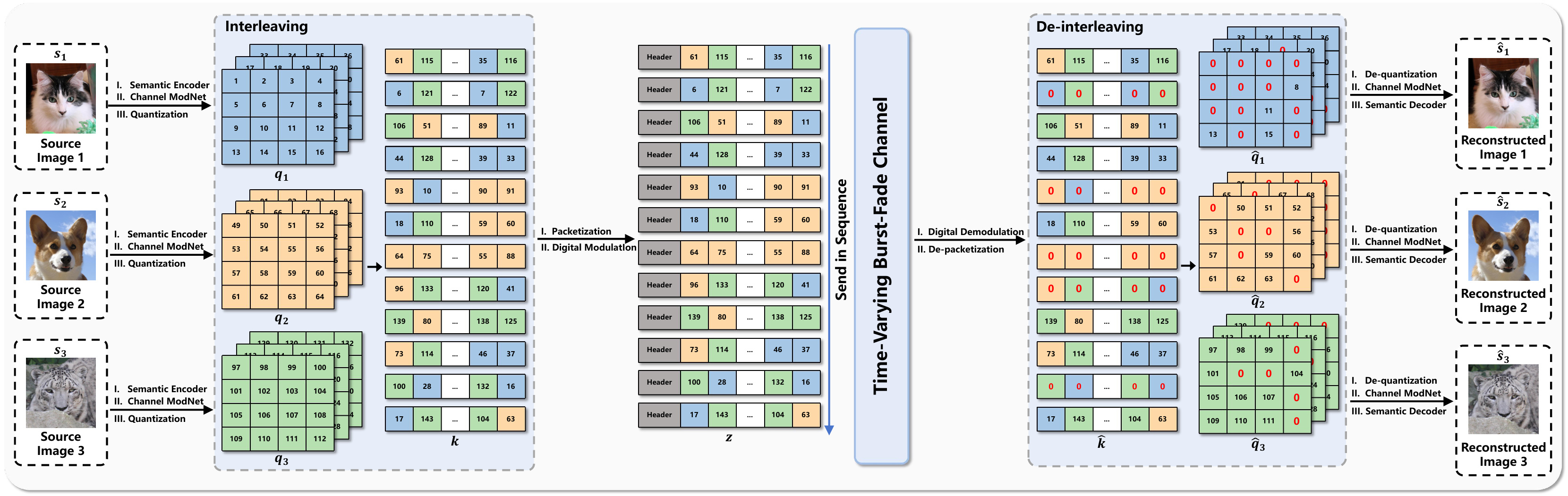}
    \caption{The cross-image semantic-level interleaving mechanism. Semantic features from multiple images are randomly interleaved before transmission to distribute correlated information across packets.}
    \label{fig:Crossimagesemanticlevelinterleaving}
\end{figure*}

Subsequently, the semantic information $\bm{x_i}$ is modulated into the semantic feature $\bm{u_i}$ by Channel ModNet with the channel state SNR. The Feature Alignment (FA) module is implemented as a sequence of three fully connected layers, where $M$ represents the dimensionality of the latent vectors. The Feature Fusion (FU) module comprises a single fully connected layer, where $C$ denotes the number of channels in the input vector $\bm{x_i}$. The overall operation can be expressed as
\begin{equation}
    \bm{u_i} = f_{\text{CM}}(\bm{x_i}, \, \text{SNR} \,;\, \bm{\Phi_{\textbf{tx}}}), \,\,\,\, \bm{u_i} \in \mathbb{R}^{M_t \times 1},
    \label{eq:CMimgs}
\end{equation}
where $f_{\text{CM}}(\, \bm{\cdot} \,;\, \bm{\Phi_{\textbf{tx}}} \,)$ represents the Channel ModNet function, parameterized by $\bm{\Phi_{\textbf{tx}}}$. 

Inspired by \cite{gong2025digital}, a quantization process is applied to discretize continuous values into integer symbols without introducing any learnable parameters. Specifically, the feature values are first normalized to the range $[0,1]$, and then a rounding operation is performed to produce discrete integer symbols. Subsequently, each integer symbol is converted into bit representation. The overall process can be formulated as
\begin{equation}
    \bm{p_i} = \left \lfloor \text{norm}(\bm{u_i}) \times (2^b-1) \right \rceil,
    \label{eq:Discrete}
\end{equation}
\begin{equation}
    \bm{q_i} = d_{b-1} d_{b-2} \cdots d_{1} d_{0}, \;\;\;\; d_{j} = \left \lfloor \frac{\bm{p_{i}}}{2^j} \right \rfloor,
    \label{eq:quanit}
\end{equation}
\begin{equation}
    \bm{q_i} = \mathcal{Q}(\bm{u_i}),
    \label{eq:quan}
\end{equation}
where $b$ denotes the number of bits per symbol, $\left \lfloor \right \rceil$ represents the rounding operation, and $d_{j}$ denotes the $j$-th bit.

The binary signal $\bm{q_i} \in \{0,1\}^{M_t \cdot 2^{b} \times 1}$ is subsequently processed by the cross-image feature-level interleaving module, which mitigates the “single-image collapse” phenomenon arising from consecutive packet losses (as detailed in Section V-C). The interleaving operation is formulated as
\begin{equation}
    \bm{k} = g_{\text{int}}(\bm{q_1}, \cdots , \bm{q_{N}} \; ; \; \bm{\Pi}), \,\,\,\, \bm{k} \in \{0,1\}^{N \cdot M_t \cdot 2^{b} \times 1}.
    \label{eq:intlvimgs}
\end{equation}

The interleaved feature $\bm{k}$ is then segmented into packets based on SITP according to the packetization process described in Section-III, which is subsequently modulated into constellation symbols for wireless transmission. The characteristics of the wireless channel are modeled in Section V-B.

The received signal undergoes digital demodulation, yielding the recovered packet $\bm{\hat{y}}$. After the multi-layer depacketization process described in Section~III, packets failing header verification are discarded, whereas verified packets are extracted to form the interleaved bitstream $\bm{\hat{k}}$. Subsequently, de-interleaving restores the original bit sequences corresponding to multiple images, expressed as
\begin{equation}
    \{ \bm{q_1}, \cdots , \bm{q_{N}} \} = g_{\text{deint}}( \bm{\hat{k}} \; ; \; \bm{\Pi}),
    \label{eq:deintlvimgs}
\end{equation}
where $g_{\text{deint}}( \cdot; \; \bm{\Pi})$ denotes the de-interleaving function. 

Subsequently, de-quantization maps the discrete integer symbols back to continuous semantic feature. The recovered features are then refined by the Channel ModNet. Finally, the Semantic Decoder reconstructs the semantic features into perceptually images. The overall process can be expressed as
\begin{equation}
    \bm{\hat{u}_i} = \mathcal{Q}^{-1}(\bm{\hat{q}_i}), \,\,\,\, \bm{\hat{u}_i} \in \mathbb{R}^{M_t \times 1},
    \label{eq:dequan}
\end{equation}
\begin{equation}
    \bm{\hat{x}_i} = f_{\text{CM}}(\bm{\hat{u}_i}, \, \text{SNR} \,;\, \bm{\Phi_{\textbf{rx}}}), \,\,\,\, \bm{\hat{x}_i} \in \mathbb{R}^{M_t \times 1},
    \label{eq:CMimgsrx}
\end{equation}
\begin{equation}
    \bm{\hat{s}_i} = f_{\text{dec}}(\bm{\hat{x}_i} \, ; \, \bm{\Theta_{\textbf{dec}}}), \,\,\,\, \bm{\hat{s}_i} \in \mathbb{R}^{H \times W \times 3}
    \label{eq:decoderimgs}
\end{equation}
where $f_{\text{CM}}(\cdot \,;\, \bm{\Phi_{\textbf{rx}}})$ represents the Channel ModNet mapping function parameterized by $\bm{\Phi_{\textbf{rx}}}$. The optimized network parameters, collectively denoted as $\{\bm{\Theta_{\textbf{enc}}}, \, \bm{\Phi_{\textbf{tx}}}, \, \bm{\Phi_{\textbf{rx}}}, \, \bm{\Theta_{\textbf{dec}}}\}$, are trained by minimizing the following objective function:
\begin{equation}
    \mathcal{L} = \frac{1}{N} \sum_{i=1}^{N} \textbf{MSE}(\bm{s_i}, \, \bm{\hat{s}_i}).
    \label{eq:loss}
\end{equation}

\subsection{Time-Varying Channel with Burst Packet Loss}

In this paper, a time-varying fading channel model is considered, where the SNR exhibits abrupt fluctuations over time. The channel initially remains in a high-SNR (good) state, then experiences a temporary deep-fade interval characterized by strong interference, and subsequently recovers to its nominal SNR level. Such temporary degradation leads to a burst packet-loss phenomenon, in which multiple consecutive packets fail to be correctly received within the fading interval.

Let the instantaneous SNR at time $t$ be denoted as $\gamma(t)$. The channel variation can then be approximated as:
\begin{equation}
    \gamma(t) = \left\{
    \begin{aligned}
         & \gamma_{\text{high}}, \;\; 0 \leq t \leq t_1, \\
         & \gamma_{\text{low}}, \;\;\; t_1 \leq t \leq t_2, \\
         & \gamma_{\text{high}}, \;\; t_2 \leq t \leq T, \\
    \end{aligned}
    \right.
    \label{eq:channel}
\end{equation}
where $\gamma_{\text{high}}$ and $\gamma_{\text{low}}$ denote the average SNR levels in the good and faded states, respectively, and $[t_1, t_2]$ represents the fading interval during which consecutive packet losses are most likely to occur. To alleviate consecutive packet losse, the cross-image feature-level interleaving mechanism introduced in Section V-C redistributes semantic features across multiple images in the temporal domain.

\subsection{Cross-image Feature Interleaving Mechanism}

To enhance robustness against burst losses caused by time-varying channel fading, a cross-image feature interleaving mechanism is introduced, as illustrated in Fig.\ref{fig:Crossimagesemanticlevelinterleaving}. Unlike conventional intra-image interleaving schemes that operate within a single frame, the proposed method redistributes semantic feature bits across multiple images in a transmission group, thereby mitigating the impact of consecutive packet losses on any single image. Let the quantized semantic feature sequence of $N$ consecutive images be denoted as
\begin{equation}
    \bm{Q} = \{ \bm{q_1}, \, \bm{q_2}, \, \cdots, \, \bm{q_N} \}, \;\;\; \bm{q_i} \in \{0,1\}^{M_t \cdot 2^{b} \times 1}.
    \label{eq:Qset}
\end{equation}

Subsequently, a semantic interleaving index vector $\bm{\Pi} \in \mathbb{N}^{N \cdot M_t \cdot 2^{b} \times 1}$ is employed to interleave semantic features across all images, as formulated in \eqref{eq:intlvimgs}. The interleaving operation randomly redistributes semantically correlated feature segments among adjacent images, ensuring that each packet carries portions of multiple images’ semantic information. At the receiver, the same interleaving index $\bm{\Pi}$ is assumed to be shared at both the transmitter and receiver. Consequently, the de-interleaving process applies the inverse operation, as described in \eqref{eq:deintlvimgs}.

\begin{figure}[t]
    \centering
    \includegraphics[width=0.48\textwidth]{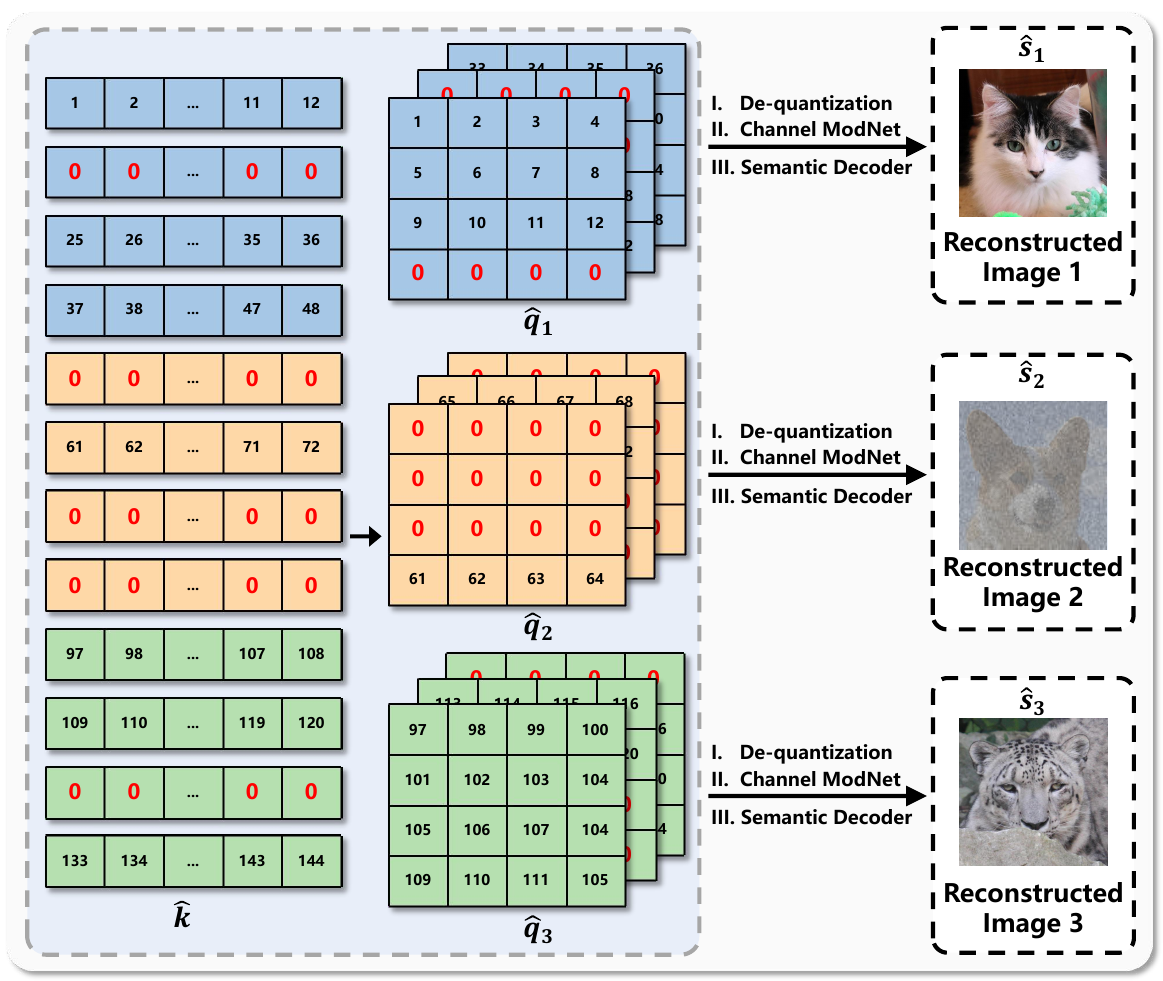}
    \caption{The semantic reconstruction without interleaving. Consecutive packet losses caused by burst fading result in concentrated semantic corruption, leading to severe degradation and missing structures in the reconstructed images.}
    \label{fig:Withoutinterleaving}
\end{figure}

Accordingly, based on the time-varying burst-fade channel model in \eqref{eq:channel}, together with the cross-layer packet loss formulation in \eqref{eq:crossfail2} and the BER model in \eqref{eq:qam}, a piecewise packet loss expression can be denoted as
\begin{equation}
    P_L(t) = \left\{
    \begin{aligned}
         & P_{\text{Cross-fail}}(P_{b}(\gamma_{\text{high}})), \;\; 0 \leq t \leq t_1 \\
         & P_{\text{Cross-fail}}(P_{b}(\gamma_{\text{low}})), \;\;\; t_1 \leq t \leq t_2 \\
         & P_{\text{Cross-fail}}(P_{b}(\gamma_{\text{high}})). \;\; t_2 \leq t \leq T \\
    \end{aligned}
    \right.
    \label{eq:channelloss}
\end{equation}

Therefore, the average packet loss probability for the image group can be expressed as follows:
\begin{align}
     \bar{P}_{L} & = \frac{t_1 \cdot P_{L}(t_1) + (t_2 - t_1) \cdot P_{L}(t_2) + (T - t_2) \cdot P_{L}(T)}{T}
     \label{eq:lossmean1} \\
     & = P_{L}(T) + \frac{(t_2 - t_1) \cdot [P_{L}(t_2)-P_{L}(T)]}{T},
     \label{eq:lossmean2}
\end{align}
where it should be noted that $T$ denotes the total number of transmitted packets within the group. Hence, the average packet loss per image group exhibits a nonlinear accumulation over time, as described in \eqref{eq:lossmean1}–\eqref{eq:lossmean2}, where the overall packet loss is jointly influenced by the duration of the degraded channel and the number of interleaved images. As the interleaving depth increases, semantic features from different images are dispersed across multiple transmission intervals, thereby mitigating burst-loss effects through probabilistic averaging. The process converts temporally concentrated burst losses into spatially distributed errors at the feature level, allowing the semantic decoder to exploit contextual redundancy for reconstruction. Consequently, the proposed cross-image feature interleaving substantially enhances robustness against burst fading.

\section{Experimental Results and Analysis}

\subsection{Experimental Setup}

% \textit{1) Datasets:} To validate the applicability of our SITP framework for image reconstruction tasks, we utilized the CIFAR-10 \cite{CIFAR10} dataset and AFHQ dataset \cite{choi2020stargan}. The CIFAR-10 dataset consists of 60,000 RGB images with the resolution of $32 \times 32$ pixels, providing a standard benchmark for low-resolution image reconstruction tasks. Futhermore, the AFHQ dataset comprises 15,000 high-quality animal face images with a resolution of $512 \times 512$. During both the training and testing phases, the images of the AFHQ dataset are resized to the dimensions of $256 \times 256$ for image reconstruction tasks.

\textit{1) Datasets:} To validate the applicability of our SITP framework for image reconstruction tasks, we utilized AFHQ dataset \cite{choi2020stargan}. The AFHQ dataset comprises 15,000 high-quality animal face images with a resolution of $512 \times 512$. During both the training and testing phases, the images are resized to the dimensions of $256 \times 256$ for image reconstruction tasks.

\textit{2) Baseline Methods:} To comprehensively evaluate the performance of the proposed SITP framework, several representative baseline methods are implemented for comparison:
\begin{itemize}
    \item {\textbf{TCP Scheme \cite{feng2025exploiting}:} TCP is adopted as a benchmark for reliable transmission. The TCP-based framework is built upon the SwinJSCC architecture extended to the digital domain, which ensures bit-level integrity through ACK and retransmission mechanisms. A maximum retransmission limit is imposed instead of assuming infinite retries.}
    \item{\textbf{UDP Scheme \cite{jiang2025position}:} UDP is considered as the baseline for low-latency communication, which performs connectionless transmission without ACK or retransmissions. Similar to the TCP configuration, the UDP-based transmission pipeline employs SwinJSCC with digital modulation.}
\end{itemize}

\textbf{Noting:} The MSTVSC method was not included as a baseline because it primarily targets video-oriented SemCom, which primarily relies on the UDP protocol. In contrast, this work focuses on the innovation of the transport-layer protocol rather than on the semantic encoder–decoder architecture.

\textit{3) Performance Metrics:} To comprehensively assess the performance of the proposed model, we adopt both semantic-level and pixel-level mertics. For semantic evaluation, the Learned Perceptual Image Patch Similarity (LPIPS) metric \cite{zhang2018unreasonable} is employed to calculate the perceptual similarity of the reconstructed images. LPIPS evaluates the perceptual difference between two images by comparing their deep feature representations extracted from a pretrained NN, with lower score indicates the higher perceptual similarity. For pixel-level evaluation, PSNR measures the fidelity of the reconstructed image, with higher values indicating reduced distortion. MS-SSIM evaluates perceptual similarity by considering luminance, contrast, and structural information across multiple scales, where higher values signify greater similarity.

\begin{figure*}[t]
    \centering
    % 第一行：AWGN Channel
    \begin{minipage}{\textwidth}
        \centering
        \includegraphics[width=0.32\textwidth]{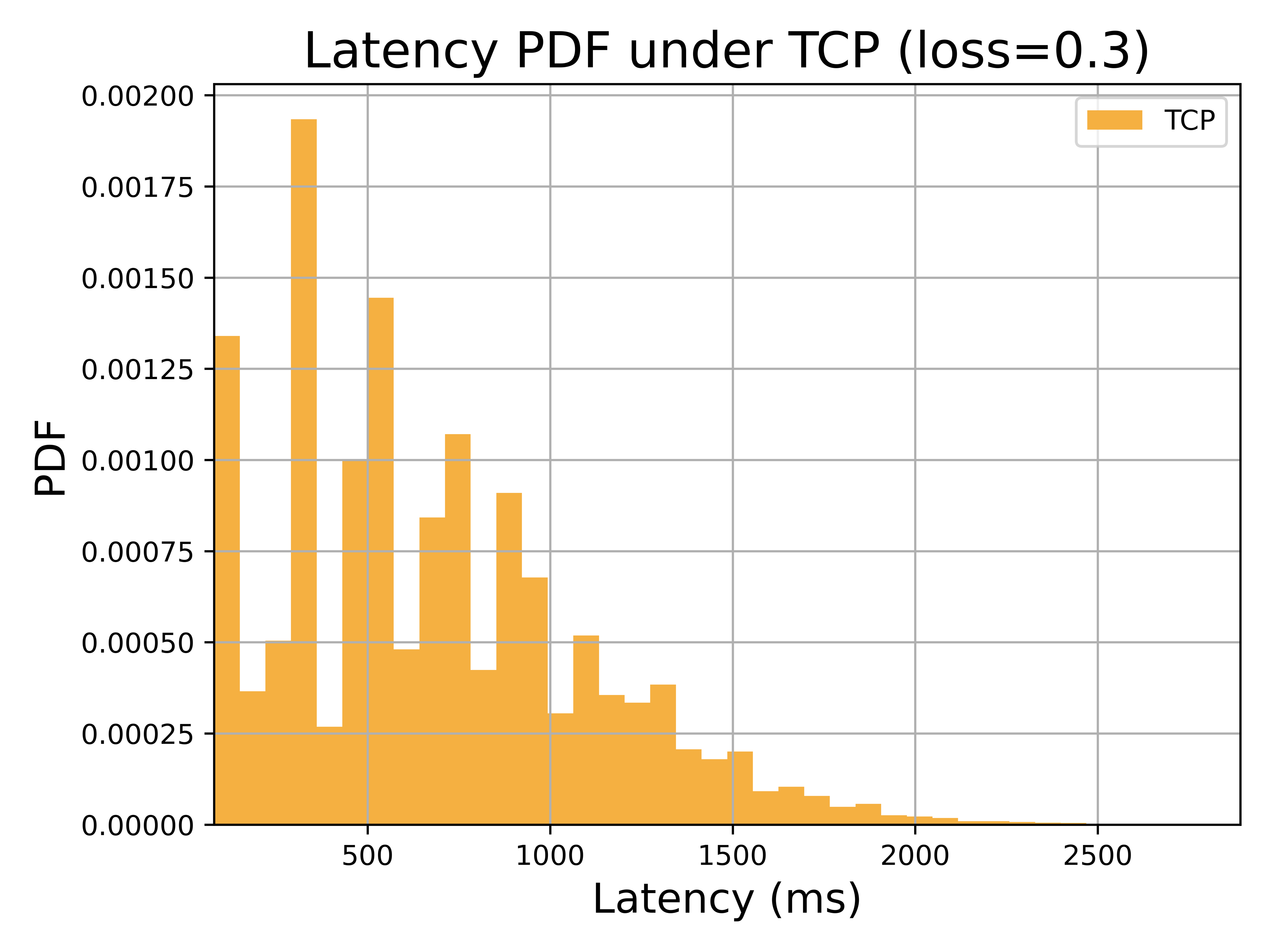} 
        \includegraphics[width=0.32\textwidth]{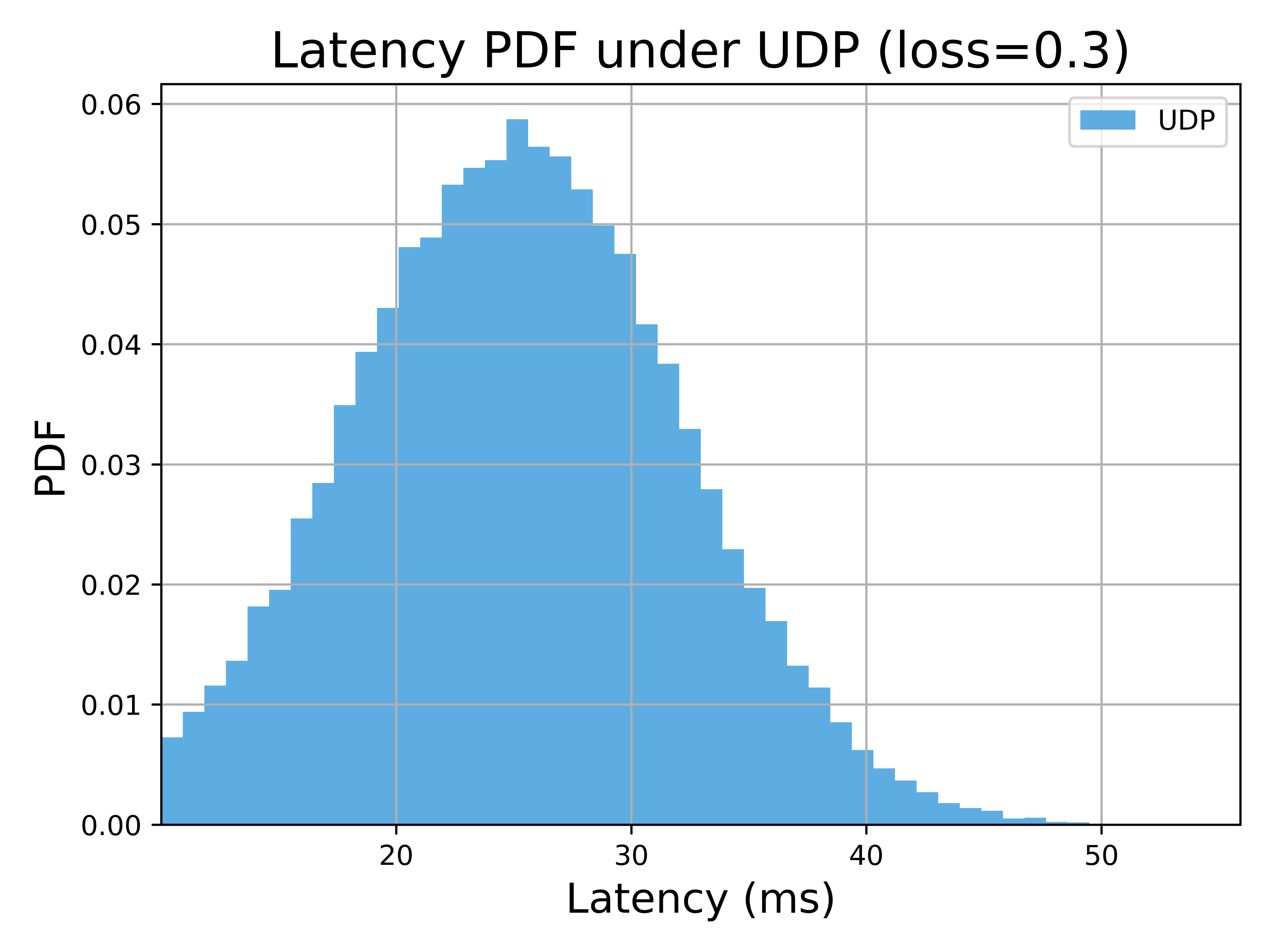} 
        \includegraphics[width=0.32\textwidth]{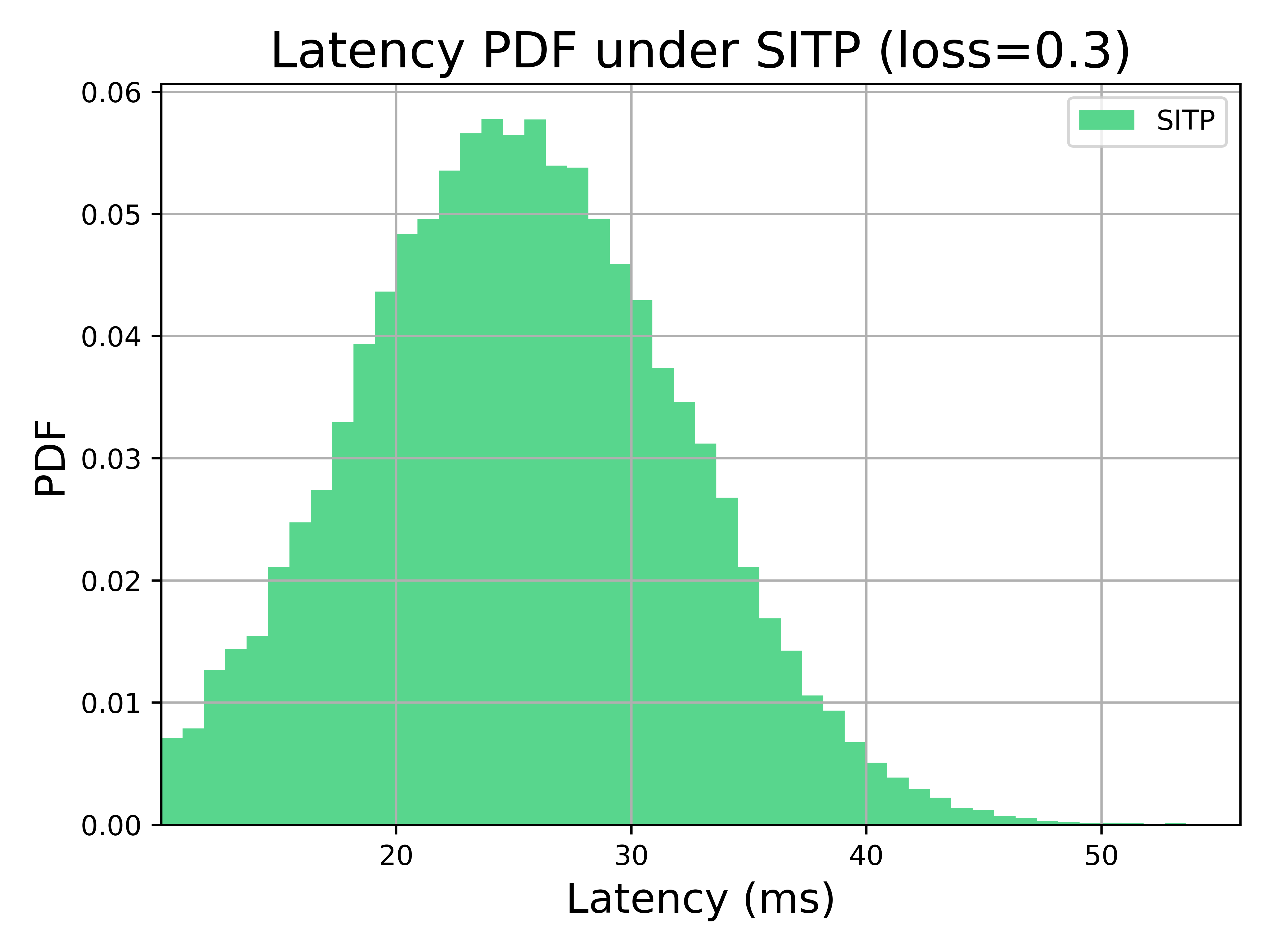} \\
        {\footnotesize (a) Latency PDFs.}
        \label{fig:latencya}
    \end{minipage}

    \vspace{2.0pt}

    % 第二行：Rayleigh Channel
    \begin{minipage}{\textwidth}
        \centering
        \includegraphics[width=0.32\textwidth]{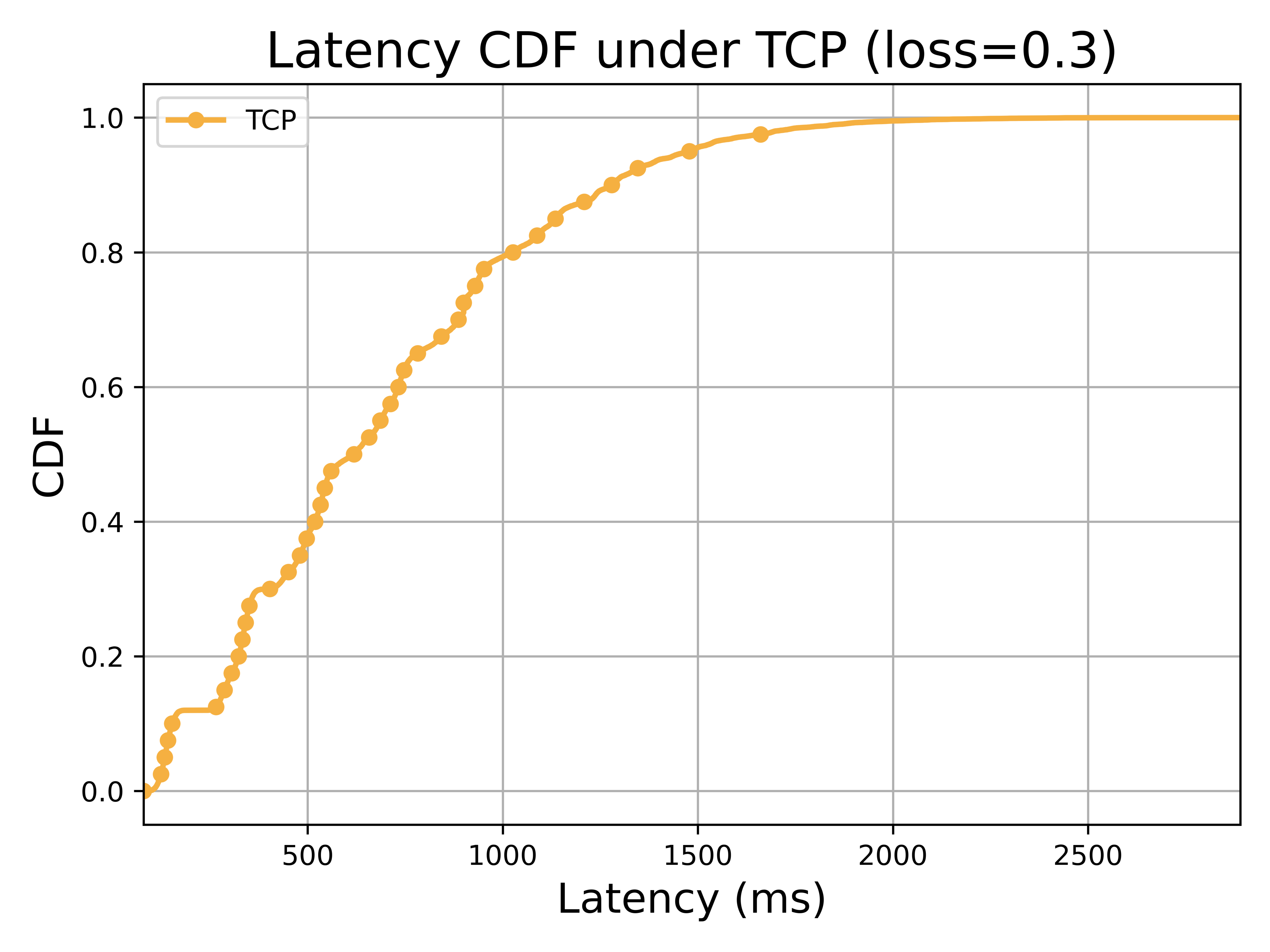}
        \includegraphics[width=0.32\textwidth]{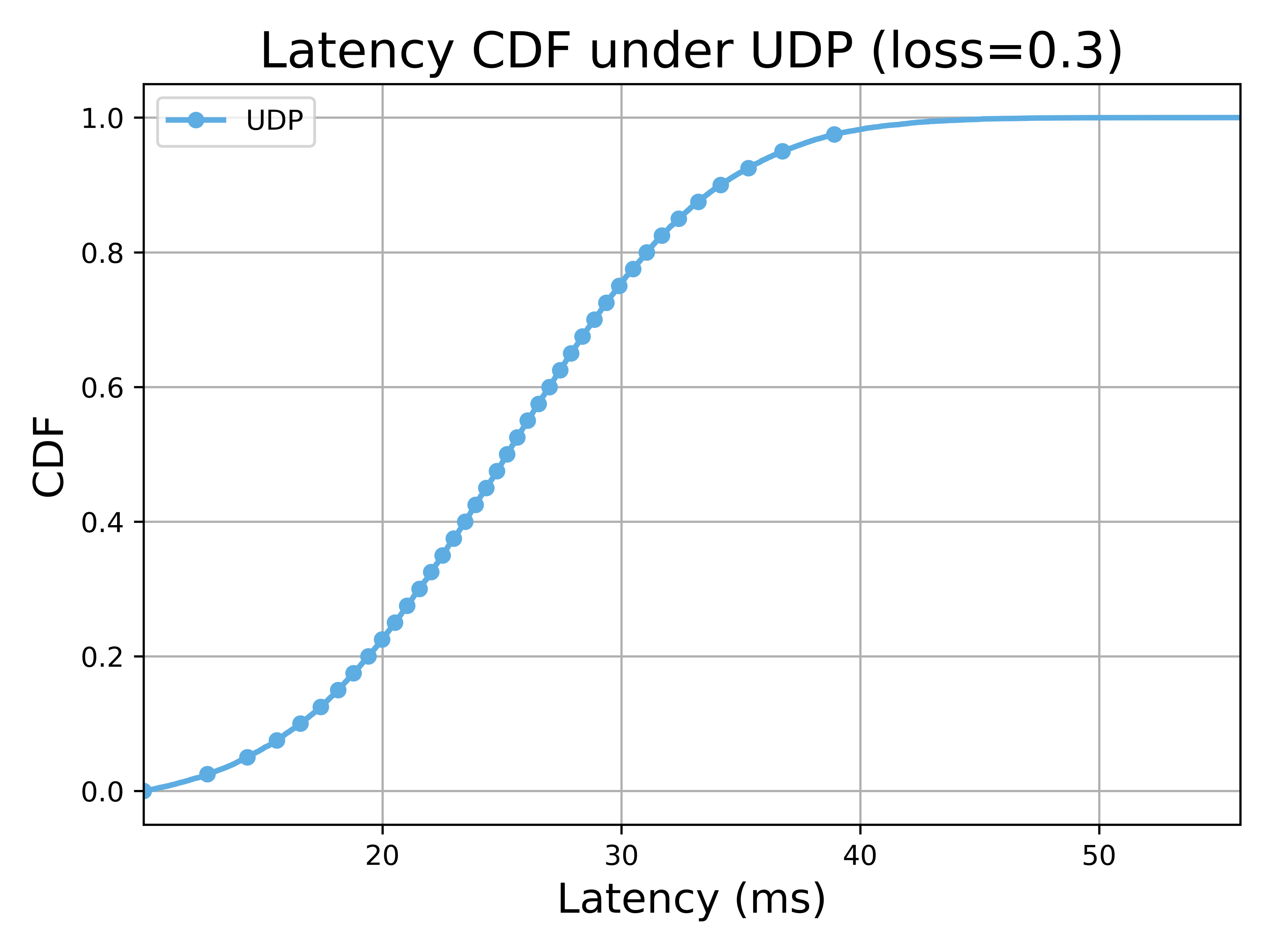}
        \includegraphics[width=0.32\textwidth]{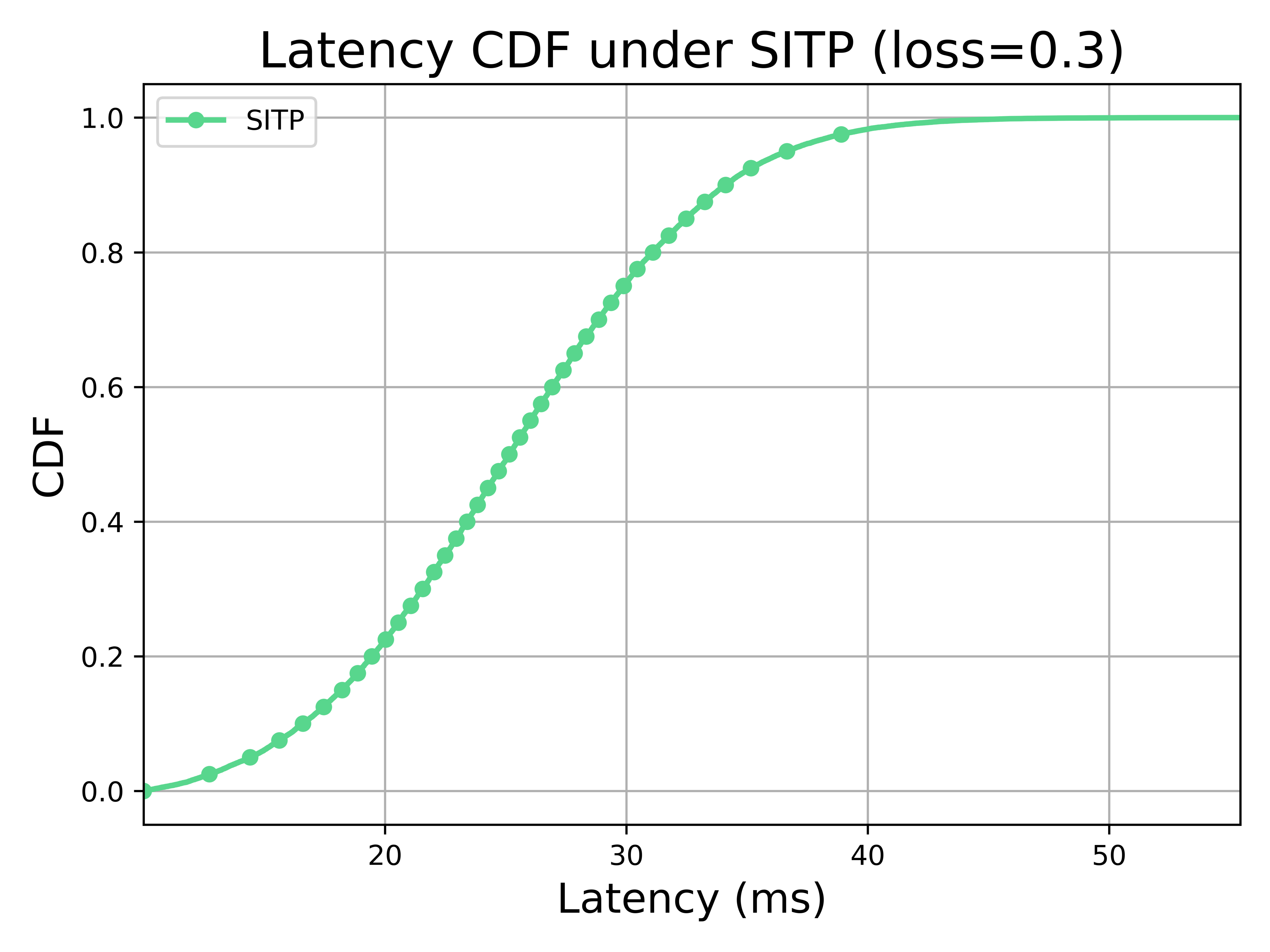} \\
        {\footnotesize (b) Latency CDFs.}
        \label{fig:latencyb}
    \end{minipage}

    \caption{Comparison of latency performance among TCP, UDP, and SITP with a packet loss rate of 0.3. (a) illustrates the latency PDFs, while (b) presents the CDFs.}
    \label{fig:latency}
\end{figure*}

\captionsetup[table]{justification=centering, labelsep=space, textfont=sc} % 设置标题全大写
\begin{table}[t]
    \renewcommand\arraystretch{1.8}
    \setlength{\tabcolsep}{4pt}
    \caption{ \\ Parameter Setting of Latency Experiment\label{tab5}}
    \centering
    \begin{tabular}{cc}
        \hline
        \hline
        Parameters & Values \\
        \hline
        Packet Loss Rate & 0.3 \\
        Monte Carlo Trials & 40000 \\
        Maximum TCP Retransmissions & 5 \\
        Mean RTT of TCP & 50.0 ms\\
        RTT Jitter of TCP & 10.0 ms \\
        Minimum RTT of TCP & 10.0 ms \\
        Mean One-way Delay of UDP & 25.0 ms \\
        Jitter of UDP & 7.07 ms \\ 
        Minimum One-way Delay of UDP & 10.0 ms \\
        \hline
        \hline
    \end{tabular}
    \label{latency}
\end{table}

\subsection{Latency Performance Analysis}

We compare the transmission latency of TCP, UDP, and the proposed SITP. For each protocol, only a single data packet is transmitted, while multiple Monte Carlo experiments are conducted to ensure statistical reliability. For the TCP protocol, both the three-way handshake and the acknowledgment (ACK) procedures are considered, and thus the round-trip time (RTT) is incorporated into the latency evaluation. In contrast, UDP and SITP perform direct transmissions without feedback or retransmission operations. To characterize the randomness of single-packet transmission latency, the impact of network jitter is explicitly considered, and the latency is modeled using a truncated Gaussian distribution. The corresponding parameters employed in the simulation are summarized in Table \ref{latency}.

Fig.\ref{fig:latency} presents the probability density functions (PDFs) and cumulative distribution functions (CDFs) of latency under a packet loss rate of 0.3. As shown in Fig.\ref{fig:latency}(a), the latency distribution of TCP exhibits a pronounced long-tail effect, indicating the presence of significant retransmission delays caused by repeated acknowledgment and timeout procedures. In contrast, both UDP and SITP demonstrate compact and symmetric latency distributions. The corresponding CDFs in Fig.\ref{fig:latency}(b) further confirm these observations. SITP achieves substantially lower end-to-end latency, primarily because it eliminates the handshake and ACK exchanges.

\begin{figure*}[t]
    \centering
    \begin{minipage}{0.32\textwidth}
        \centering
        \includegraphics[width=\textwidth]{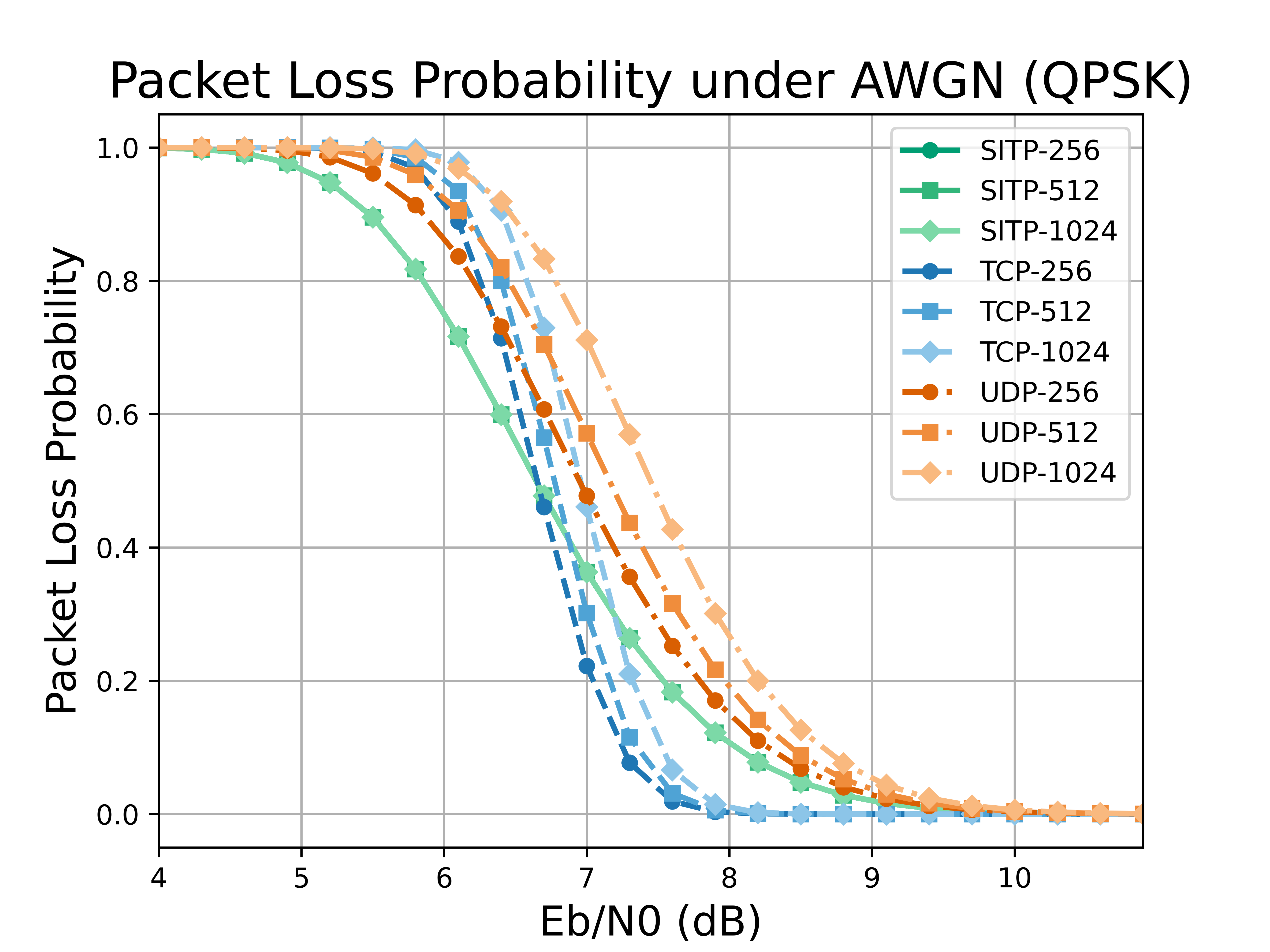} \\
        {\footnotesize (a) QPSK}
        \label{fig:snr_pra}
    \end{minipage}
    \hfill
    \begin{minipage}{0.32\textwidth}
        \centering
        \includegraphics[width=\textwidth]{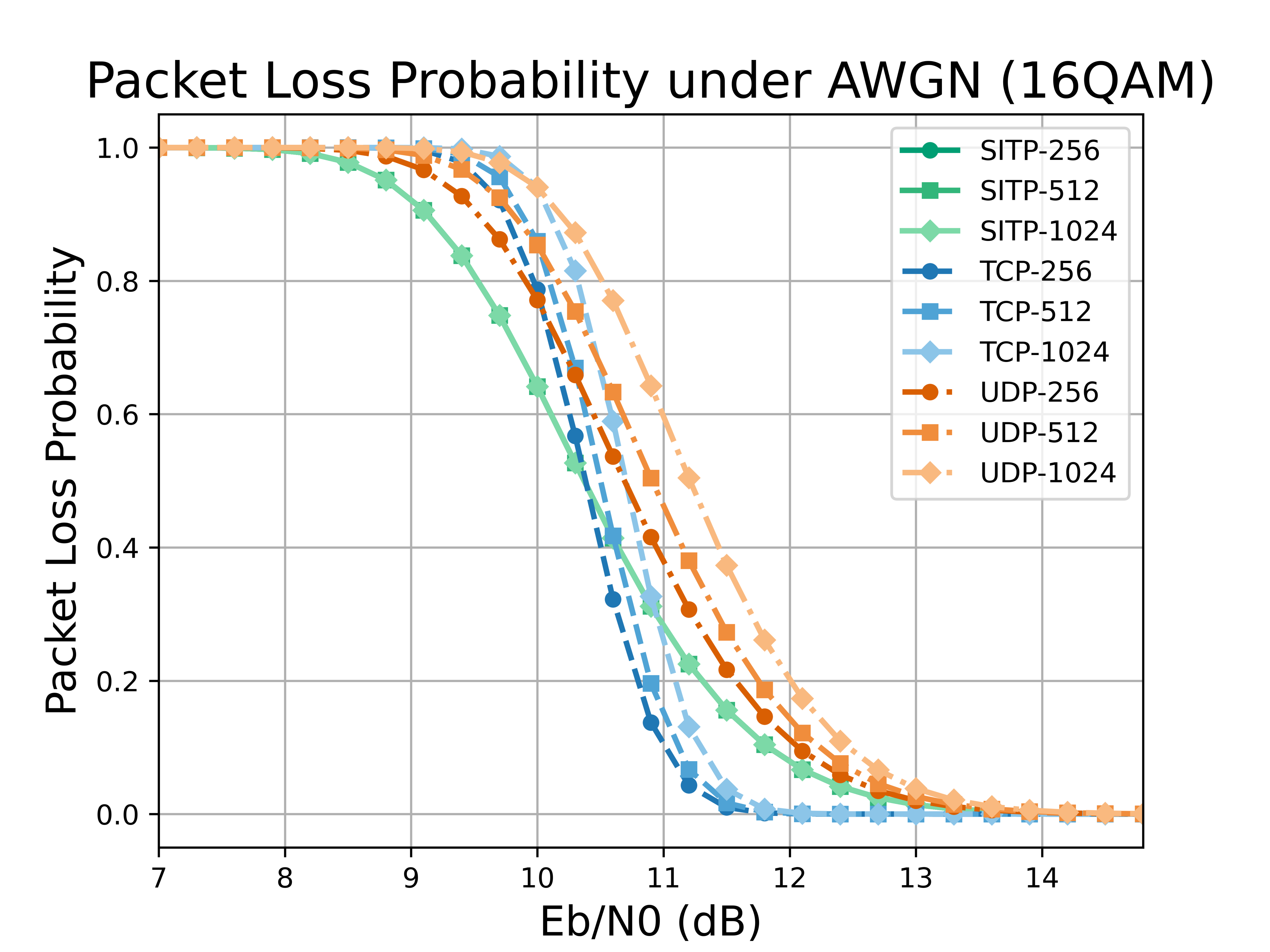} \\
        {\footnotesize (b) 16QAM}
        \label{fig:snr_prb}
    \end{minipage}
    \hfill
    \begin{minipage}{0.32\textwidth}
        \centering
        \includegraphics[width=\textwidth]{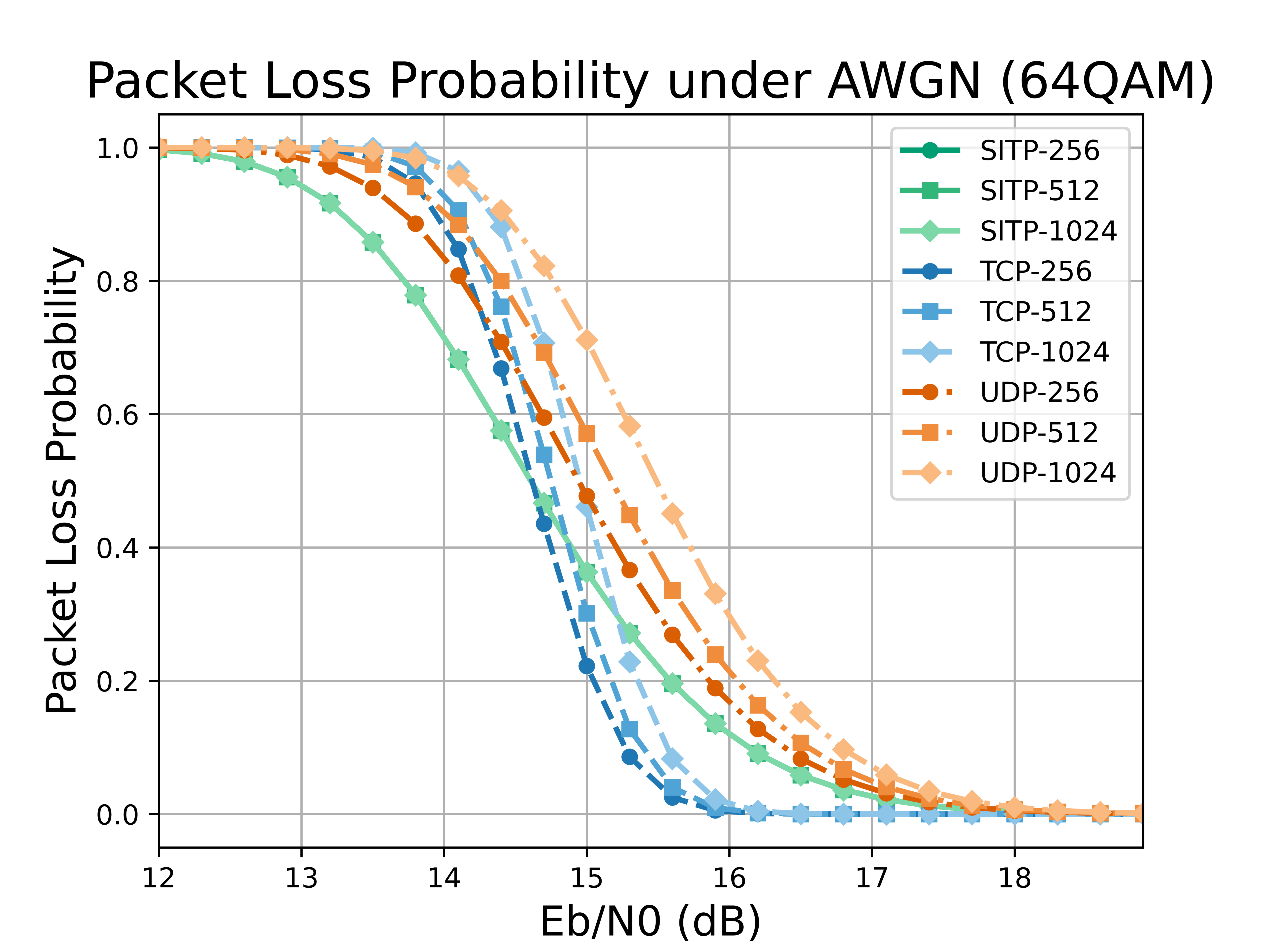} \\
        {\footnotesize (c) 64QAM}
        \label{fig:snr_prc}
    \end{minipage}
    \caption{Packet loss probability versus $E_b/N_0$ over AWGN channels for different modulation schemes (QPSK, 16QAM, 64QAM). Note: The numeric values following each protocol (e.g., 256, 512, 1024) denote the payload length $L$ of the data segment within each packet.}
    \label{fig:snr_pr}
\end{figure*}

\begin{figure*}[t]
    \centering
    \begin{minipage}{0.32\textwidth}
        \centering
        \includegraphics[width=\textwidth]{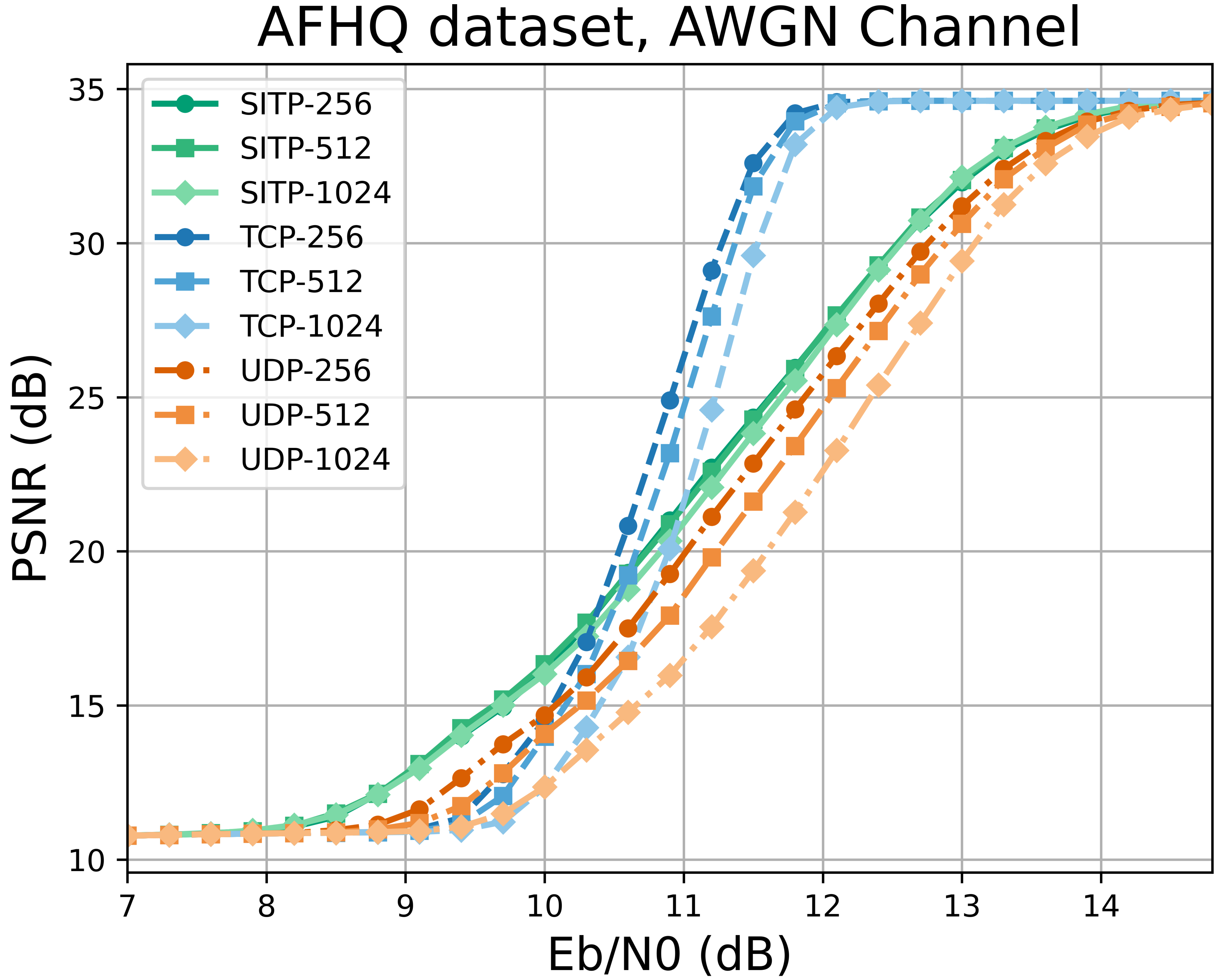} \\
        {\footnotesize (a) PSNR}
        \label{fig:snr_performancea}
    \end{minipage}
    \hfill
    \begin{minipage}{0.32\textwidth}
        \centering
        \includegraphics[width=\textwidth]{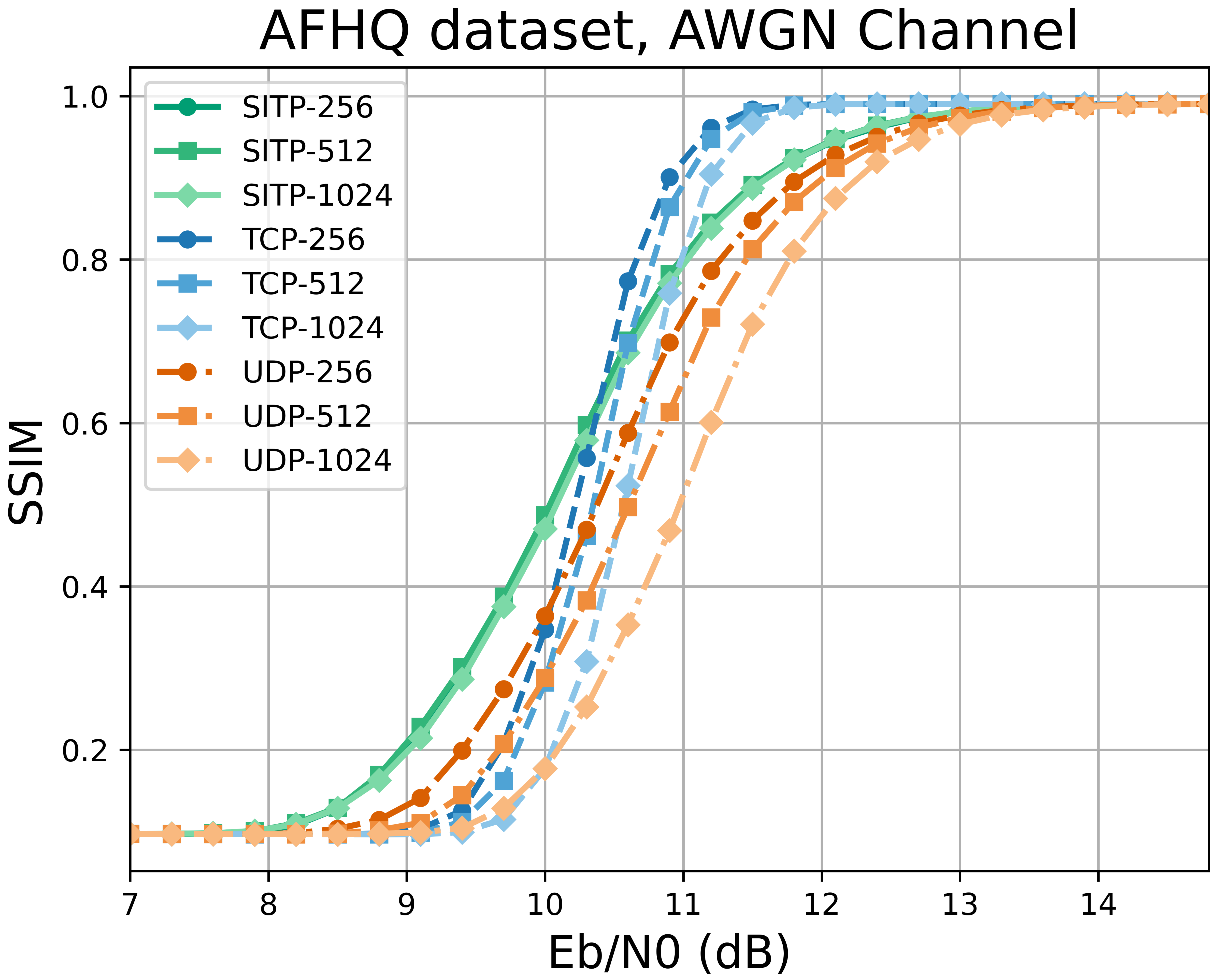} \\
        {\footnotesize (b) MS-SSIM}
        \label{fig:snr_performanceb}
    \end{minipage}
    \hfill
    \begin{minipage}{0.32\textwidth}
        \centering
        \includegraphics[width=\textwidth]{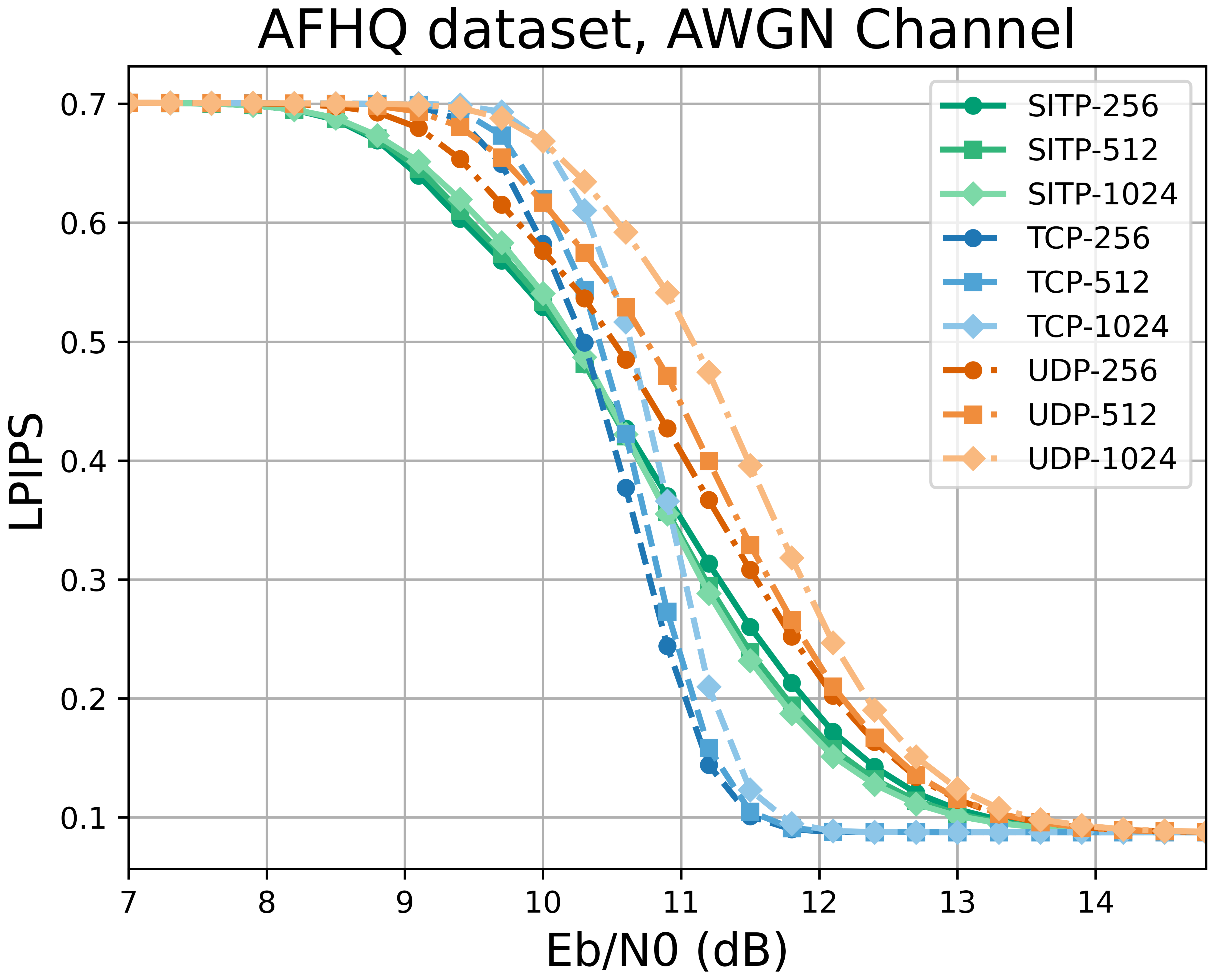} \\
        {\footnotesize (c) LPIPS}
        \label{fig:snr_performancec}
    \end{minipage}
    \caption{Performance comparison among TCP, UDP, and SITP over AWGN channels on the AFHQ dataset. The results indicate that SITP achieves significantly higher PSNR and MS-SSIM values, and lower LPIPS scores than TCP at low SNRs,, while maintaining comparable performance to UDP across all payload lengths.}
    \label{fig:snr_performance}
\end{figure*}

\captionsetup[table]{justification=centering, labelsep=space, textfont=sc} % 设置标题全大写
\begin{table}[t]
    \renewcommand\arraystretch{1.8}
    \setlength{\tabcolsep}{4pt}
    \caption{ \\ Parameter Setting of Packet Loss Experiment\label{tab5}}
    \centering
    \begin{tabular}{cccc}
        \hline
        \hline
        Parameters & Values (bits) & Parameters & Values (bits) \\
        \hline
        $N_{\text{PH}}$ & 64 & $N_{\text{sync}}$ & 11 \\
        $t_{\text{sync}}$ & 3 & $N_{\text{DH}}$ & 112 \\
        $r_{d}$ & 32 & $N_{\text{NH}}$ & 320 \\
        $N_{\text{SITP\_HDR}}$ & 64 & $N_{\text{UDP\_HDR}}$ & 64 \\
        $N_{\text{TCP\_HDR}}$ & 224 & $r_{s}$ & 16 \\
        $N_{\text{AH}}$ & 24 & $L$ & [256, 512, 1024] \\
        \multicolumn{2}{c}{Protocols} & \multicolumn{2}{c}{[TCP, UDP, SITP]} \\
        \multicolumn{2}{c}{Digital Modulation} & \multicolumn{2}{c}{[QPSK, 16QAM, 64QAM]} \\
        \multicolumn{2}{c}{Maximum TCP Retransmissions} & \multicolumn{2}{c}{5} \\
        \hline
        \hline
    \end{tabular}
    \label{packetloss}
\end{table}

\subsection{Packet Loss Performance Validation}

To evaluate the robustness of the proposed SITP framework, we conducted a comparative packet loss analysis against TCP and UDP under varying SNRs. For the TCP simulations, a limited number of retransmissions was considered to better reflect practical scenarios. The relationship between SNR and BER was modeled according to the analytical formulation given in \eqref{eq:qam}. To further assess the adaptability of SITP to different transmission conditions, multiple payload lengths $L$ were tested to account for diverse data segment sizes. The detailed parameter configurations are summarized in Table \ref{packetloss}.

Fig.\ref{fig:snr_pr} illustrates the packet loss probability models under different modulation schemes. As observed, the packet loss rate of SITP remains almost unaffected by the payload length $L$, since checksum verification is applied exclusively to the packet header, whereas the payload portion is accepted even when partially corrupted. Consequently, the packet loss probability of SITP is independent of the data segment length, providing greater flexibility in packet assembly. In contrast, both TCP and UDP validate the entire payload during transmission, resulting in higher packet loss probabilities as $L$ increases. However, TCP benefits from its retransmission mechanism, which partially mitigates this degradation. Overall, SITP consistently achieves lower packet loss probabilities than UDP across the entire SNR range and outperforms TCP in low-SNR regimes, demonstrating superior reliability.

\subsection{Reliability Performance Analysis}

We employ an image-oriented SemCom framework to validate the reliability advantages of the proposed SITP system through end-to-end training. The semantic transceiver is developed based on the SwinJSCC architecture and further extended to accommodate digital communication scenarios. The simulation parameters are summarized in Table \ref{packetloss}. Different from the previous setup, only 16QAM is adopted as the digital modulation scheme in this experiment.

As shown in Fig.\ref{fig:snr_performance}, the proposed SITP framework exhibits superior reliability performance compared with both TCP and UDP across different SNR regimes. Specifically, SITP achieves noticeably higher PSNR and MS-SSIM values and lower LPIPS scores than TCP in low-SNR conditions, while consistently outperforming UDP over the entire SNR range, which is consistent with the packet-loss behavior observed in Fig.\ref{fig:snr_pr}, thereby validating the theoretical and empirical consistency of the proposed system. Furthermore, the results indicate that the performance of SITP still remains largely unaffected by the payload length $L$.

\begin{figure*}[t]
    \centering
    \begin{minipage}{0.32\textwidth}
        \centering
        \includegraphics[width=\textwidth]{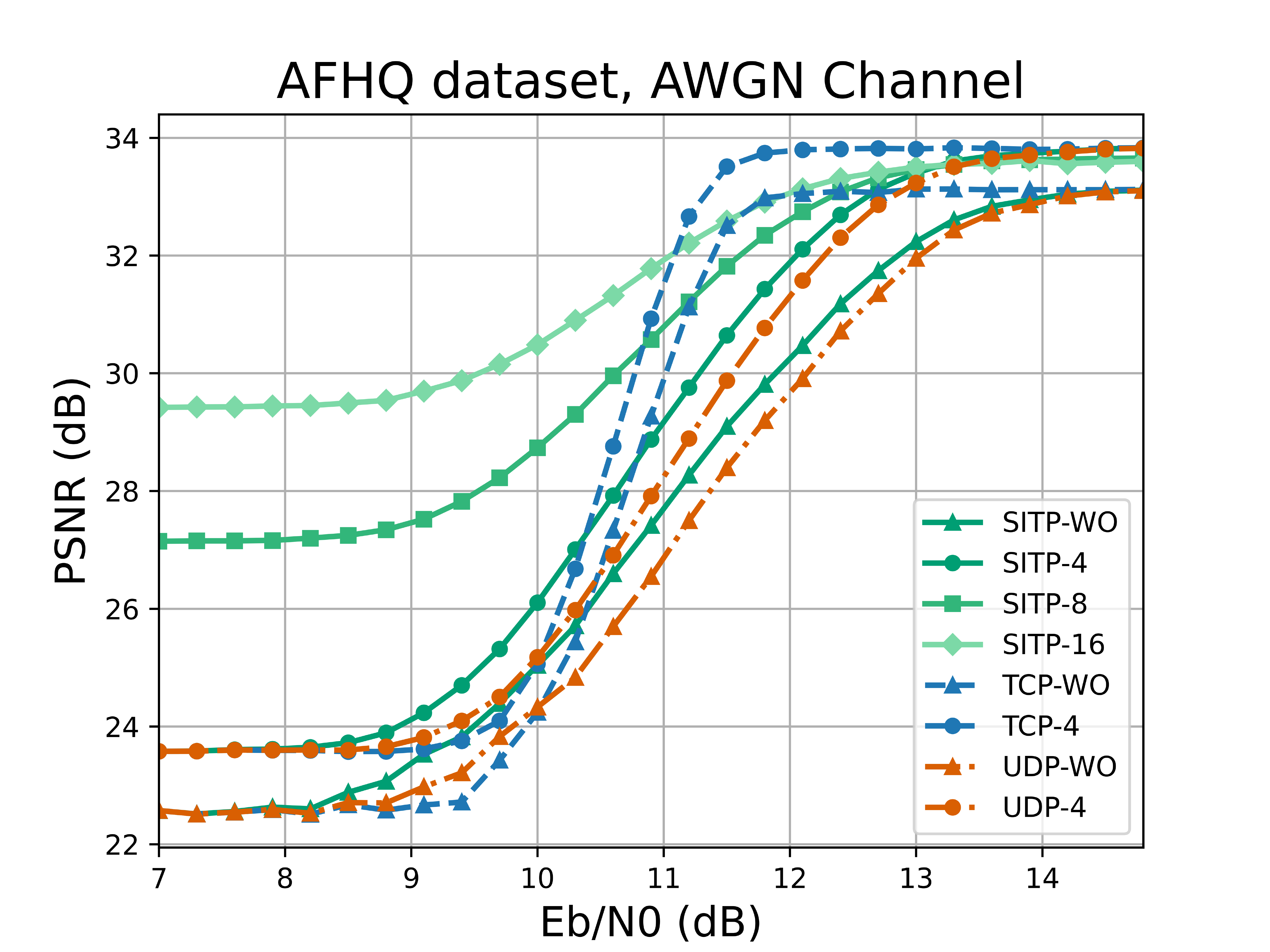} \\
        {\footnotesize (a) PSNR}
        \label{fig:snr_intera}
    \end{minipage}
    \hfill
    \begin{minipage}{0.32\textwidth}
        \centering
        \includegraphics[width=\textwidth]{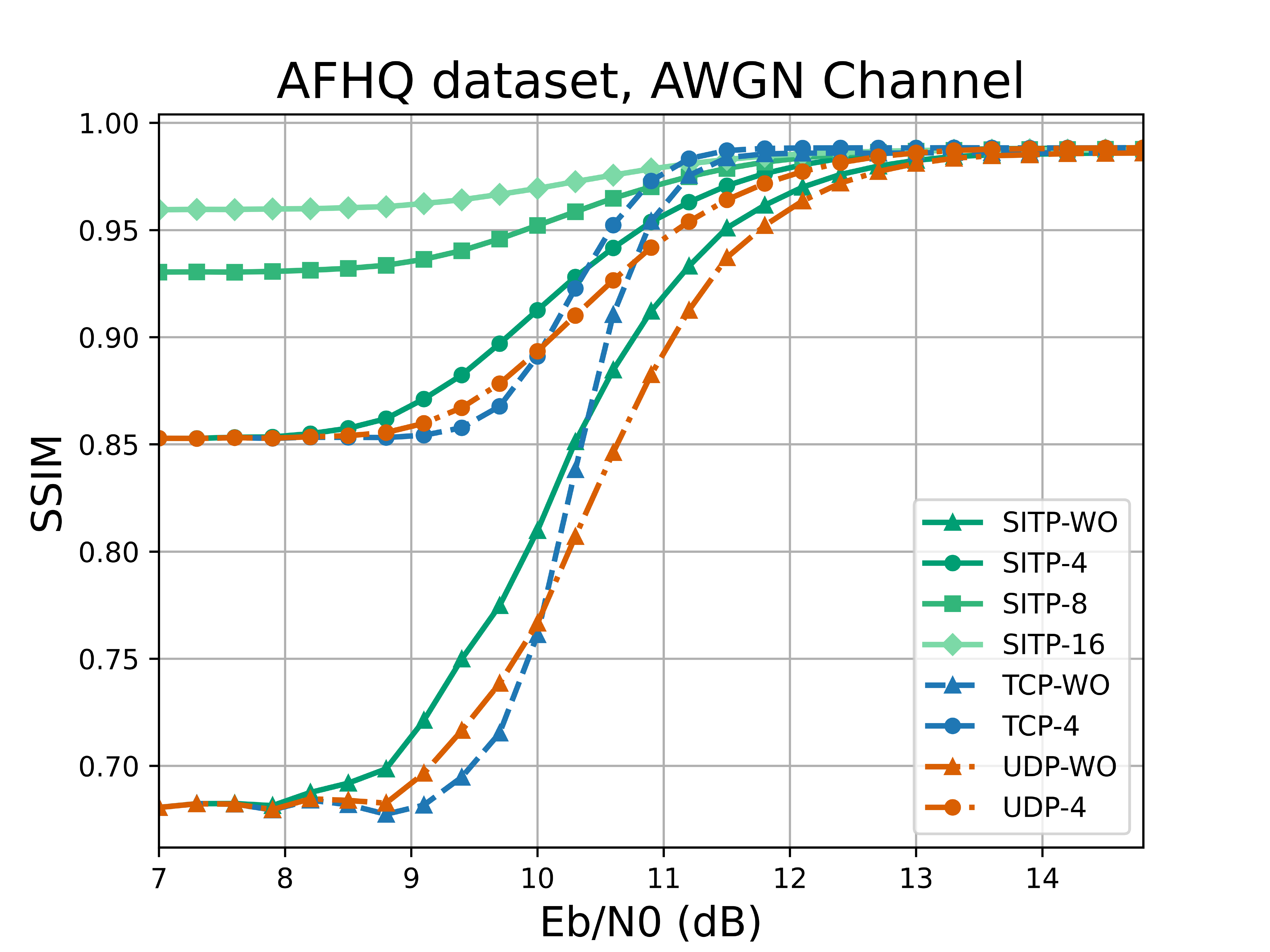} \\
        {\footnotesize (b) MS-SSIM}
        \label{fig:snr_interb}
    \end{minipage}
    \hfill
    \begin{minipage}{0.32\textwidth}
        \centering
        \includegraphics[width=\textwidth]{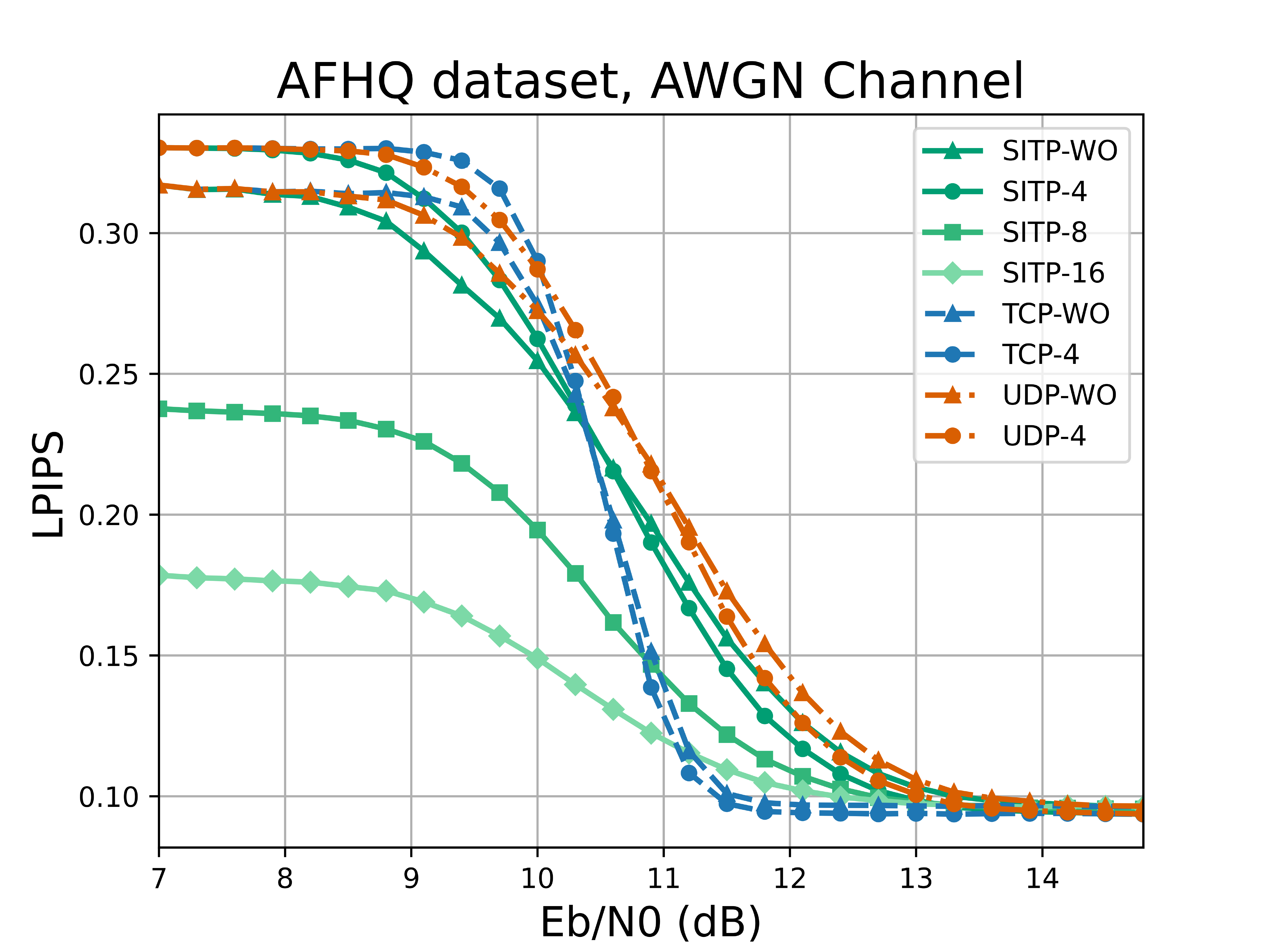} \\
        {\footnotesize (c) LPIPS}
        \label{fig:snr_interc}
    \end{minipage}
    \caption{Performance comparison of the proposed SITP framework with TCP and UDP under AWGN channels on the AFHQ dataset, with and without interleaving.}
    \label{fig:snr_inter}
\end{figure*}

\captionsetup[table]{justification=centering, labelsep=space, textfont=sc} % 设置标题全大写
\begin{table}[t]
    \renewcommand\arraystretch{1.8}
    \setlength{\tabcolsep}{4pt}
    \caption{ \\ Parameter Setting of Interleaving Experiment\label{tab5}}
    \centering
    \begin{tabular}{cccc}
        \hline
        \hline
        Parameters & Values & Parameters & Values \\
        \hline
        $t_2 - t_1$ & 528 & $\gamma_{\text{good}}$ & $15$ dB  \\
        batch size & [4, 8, 16] & $L$ & $256$ bits \\
        \multicolumn{2}{c}{$\gamma_{\text{bad}}$} & \multicolumn{2}{c}{[7.0, 7.3, $\cdots$, 14.5, 14.8] dB} \\
        \multicolumn{2}{c}{Channel Type} & \multicolumn{2}{c}{Time-Varying Burst-Fade Channel} \\
        \hline
        \hline
    \end{tabular}
    \label{Interleavingsetting}
\end{table}

\subsection{Interleaving Performance Analysis}

To evaluate the effectiveness of the proposed interleaving mechanism, comparative experiments were conducted under TCP, UDP, and SITP transmission schemes, with and without interleaving. Different interleaving depths, corresponding to batch sizes of [4, 8, 16], were selected to verify the analytical correctness of \eqref{eq:lossmean2}. Since non-interleaved transmission is independent of interleaving depth, its batch size was fixed at 4 for consistency. In the time-varying burst-fade channel model defined in \eqref{eq:channelloss}, the duration of the degraded channel state $t_2-t_1$ was set to the equivalent of 528 packet transmissions, while the start time $t_1$ was randomly initialized within a sliding window. The detailed simulation parameters are summarized in Table \ref{Interleavingsetting}.

As illustrated in Fig.\ref{fig:snr_inter}, the proposed cross-image interleaving mechanism significantly enhances transmission robustness under burst-fade channel conditions. With increasing interleaving depth (batch sizes of 4, 8, and 16), the impact of burst packet losses is effectively mitigated, as the corrupted packets are redistributed across multiple images rather than concentrated within a single frame. Furthermore, it can be observed that the proposed cross-images feature-level interleaving mechanism is compatible with both TCP and UDP frameworks, demonstrating its universality and potential applicability across diverse transport-layer protocols.

\section{Conclusion}

A novel transport-layer framework termed the Semantic Information Transport Protocol (SITP) is proposed to meet the high reliability and low latency of SemCom systems. The protocol shifts the validation focus from bit-level accuracy to header-only verification, departing from the retransmission mechanism of TCP and the discard-on-error behavior of UDP. Furthermore, a cross-layer packet-loss model is developed to capture the interactions among the physical, data-link, network, transport, and application layers within a unified analytical framework, which offers a theoretical foundation. In addition, a cross-image semantic interleaving strategy is introduced to enhance robustness under burst-fade conditions by distributing semantic features across multiple correlated frames, effectively mitigating the effects of consecutive packet losses. Experimental results confirm that SITP achieves performance comparable to TCP in reliability and to UDP in latency, while providing higher semantic reconstruction quality.

While the SITP framework has achieved promising performance across image datasets, the present work has been confined to image transmission. Future research will extend the framework to support multi-user SemCom diverse data modalities such as text, speech, and video.

\end{document}